\documentclass[useAMS,usenatbib]{mn2e}
\usepackage{graphicx}

\title[Supersolar metallicity in G0--G3 stars] {Supersolar metallicity in G0--G3 main sequence 
stars  with V$<$15}
\author[R. L{\'o}pez-Valdivia et al.]{R. L{\'o}pez-Valdivia$^{1}$\thanks{E-mail:
valdivia@inaoep.mx}, E. Bertone$^{1}$, M. Ch{\'a}vez$^{1}$, C. Tapia-Schiavon$^{1,2}$,\newauthor 
J. B. Hern\'andez-\'Aguila$^{1}$, J. R. Vald\'es$^{1}$, and V. Chavushyan$^{1}$\\
$^{1}$Instituto Nacional de Astrof{\'i}sica, {\'O}ptica y Electr{\'o}nica, Luis Enrique Erro 1, Tonantzintla, 
Puebla, 72840, M{\'e}xico \\
$^{2}$Centro de Radioastronom{\'i}a y Astrof{\'i}sica, Universidad Nacional Aut{\'o}noma de M{\'e}xico, 
Apdo. Postal 3–72 (Xangari),\\
58090 Morelia, Michoac{\'a}n, M{\'e}xico}

\begin{document}

\date{}

\pagerange{\pageref{firstpage}--\pageref{lastpage}} \pubyear{2002}

\maketitle

\label{firstpage}

\begin{abstract}
The basic stellar atmospheric parameters (effective temperature, 
surface gravity and global metallicity) were simultaneously determined 
for a sample of 233 stars, limited in magnitude ($V$~$<$15) with spectral types between G0 and G3 and 
luminosity class V (main sequence). The analysis was based on spectroscopic 
observations collected at the Observatorio Astrof\'isico Guillermo Haro and using a set of 
Lick-like indices  defined in the spectral range of 3800--4800~\AA. 
An extensive set of indices computed in a grid of theoretical spectra was used as a 
comparison tool in order to determine the photospheric parameters.
The method was validated by matching the results 
from spectra of the asteroids Vesta and Ceres with the Sun parameters. The main results were: 
i) the photospheric parameters were determined for the first time for 213 objects in our sample; 
ii) a sample of 20 new super metal-rich stars candidates was found. 

\end{abstract}


\section{Introduction}

Stellar atmospheric parameters are of fundamental importance in a plethora
of astrophysical scenarios. On the one hand, effective temperature and
surface gravity ($T_{\rm eff}$ and $\log{g}$) and, to some extent, chemical
composition allow, for instance, to properly locate stars
in the HR diagram and therefore to better establish their evolutionary
status and ages. This has been particularly true for segregating
stars on the main sequence from evolved objects in the {\it Kepler} field in
a wide temperature range \citep{molenda13}. They are
also important in identifying solar analogues or twins that permit to
place our Sun into context of its neighbourhood \citep{porto14}.\\
On the other hand, the stellar chemical composition turns out to be a
primary criterion in the analysis of the chemical history of galaxies,
including our Milky Way. Global metallicity ([M/H]) provides, in the case of metal-poor
stars, information on the early stages of galaxy evolution, before the rapid
insertion of enriched yields through supernovae events shaped their
observable chemical properties. Less attention, however, was initially paid
to the super metal-rich (SMR) stars, in spite of having been defined more than
four decades ago by \cite{spinrad69} as stars with metallicity higher than the Hyades. This leading (and
controversial) study and that of \cite{rich88} on the galactic bulge have for
a long time served as the basis for investigations of metal-rich
extragalactic populations, such as those found in elliptical galaxies and
spheroidal components (mainly bulges) of spirals.
In this work, we adopt a threshold of +0.16~dex for a star to be considered SMR; this value is slightly more conservative than the recent estimate of the average metallicity of the Hyades
([Fe/H]=0.13$\pm$0.06~dex), reported by \citet{heiter14} \\
More locally, SMR stars have become particularly interesting objects
in view of the well established correlation between the
presence of giant exoplanets and the stellar metal content \citep{gonzalez98,
santos01,fisher05}. Such correlation indicates that
the probablility of finding a giant planet significantly increases with
increasing metallicity. It is now known that 25\% of metal-rich nearby 
field stars harbor planets, while the prevalence is reduced to 3\% if we consider solar
abundances \citep{santos08}. The metallicity-planet correlation has
posed interesting challenges to the current planet formation scenarios,
since it appears to favour the core accretion model \citep{alibert2004}
over the planetary formation as a result of disk instabilities \citep{boss97}.
It has also motivated a number of studies in search for giant exoplanets in
field stars as well as in metal-rich stellar clusters as NGC~6791 \citep{bruntt03}.\\

To date, SMR stars are attractive on both of the above scenarios, 
and studies in one field have been shared by the other. 
For instance, the construction of stellar databases as tools for the synthesis 
of stellar population have recently been incorporated in exoplanetary 
studies \citep{buzzoni01}. Conversely, stellar studies, aimed at finding fiducial 
targets for exoplanet searches and potential correlations between stellar host 
properties and the presence of substellar companions, have already impacted 
broader issues, as the chemical evolution of the Milky Way \citep[see][and 
references therein]{neves09,adibekyan11,adibekyan12}.

In this work, we carried out a spectroscopic analysis of a sample of 
stars of spectral types G0--G3, 
luminosity class V, in the visual magnitude interval $4.05<V<14.77$, that has 
been observed at moderate resolution
(FWHM~2.5~\AA). We simultaneously derive the three leading atmospheric
parameters ($T_{\rm eff}$, $\log{g}$ and [M/H]), through the study of
their spectroscopic indices, and identify a new sample of SMR stars as
potential targets of planet searches.

\section[]{Stellar sample and observations}
\begin{figure}
\includegraphics[width=85mm]{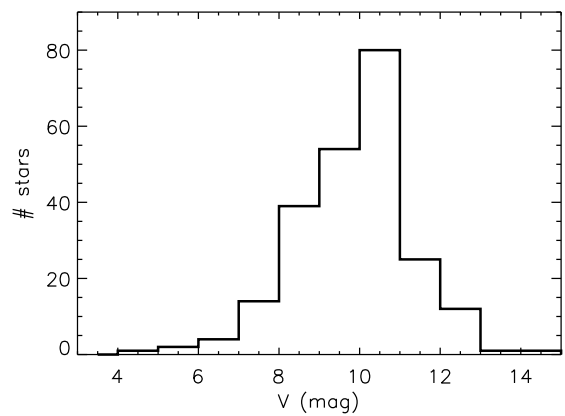}
\caption{Distribution of V magnitude for all the sample.}
\label{fig:vdist}
\end{figure}

We selected in late 2008 a sample of stars from  SIMBAD database using the following criteria: {\it i)} spectral type between G0 and G3; {\it ii)} luminosity 
class V; {\it iii)} visible magnitude $V$~$<$15~mag; and {\it iv)} declination $\delta > -10\degr$. The selection 
resulted in about 1200 objects. We report here the results for 233 stars. 
We carried out the spectroscopic observations at the 2.12 meter telescope of the Observatorio 
Astrof\'isico Guillermo Haro (Sonora, Mexico) between 2008 and 2013, with a Boller \& Chivens 
spectrograph, equipped with a Versarray 1300$\times$1340 CCD. The grating of 600 l/mm and the slit width of 
200~$\mu$m provided a constant spectral resolution of 2.5~\AA\ FWHM and a dispersion of 0.7~\AA~px$^{-1}$ along the 
wavelength range between about 3800 and 4800~\AA.
Typically, we acquired 2 or more images for each star, 
with total exposure times between 3~min, for the brighter objects, and about 60~min for the fainter ones, to 
reach a signal-to-noise ratio per pixel (S/N) of 30--80, computed in a window of 100~\AA\ around 
4600~\AA\footnote{The S/N is the average value given by a moving standard deviation over a 5~\AA-wide window. Since 
in the spectra part of the standard deviation is caused by real absorption lines, the S/N values, which therefore 
also depend on photometric parameters, should be considered as lower limits.}. The sample is presented in 
Table~\ref{tab:samp}, where we list the star identification, the $V$ magnitude and the spectral type as 
given in SIMBAD\footnote{Since the first selection, back in 2008, six stars no longer would 
accomplish the spectral type criterion. Three objects (HD~19373, HD~100180, and BD+31~4162) 
are still classified as main sequence and their re-classification ascribes them a slightly 
hotter status of, at most, one spectral sub-class. The star HD~105898 appears originally 
classified as G2~V, in Table VII of \citet{eggen64}, but now the classification 
corresponds to a cooler and evolved object (G6~III-IV). HD~137510 appears classified as G2~V in 
\citet{harlan70}, and currently the SIMBAD type is G0~IV-V. Finally,
the star HD~149996 now lacks luminosity class, although it has been designated as dwarf
by \citet{bonifacio00}.}. In Fig.~\ref{fig:vdist} we show the distribution of the $V$ magnitude for the stars in our sample.

In addition to the sample described above, we also observed a set of 29 
\textit{reference} stars from the PASTEL catalogue \citep{pastel} and complemented this sample with 35 objects from 
\citet{liu08}.
The reference sample will serve to test the adequacy of theoretical spectra in a range of atmospheric parameters, 
as explained in Sect.~\ref{sec:calib}. The main photospheric parameters (see Table~\ref{tab:refs}) of these objects were 
obtained by averaging only those PASTEL entries that provide the three parameters 
($T_{\rm eff}$,$\log{g}$,[M/H]) to minimize the effect of degeneracies.  
The reference stars span 4857--6987~K in effective temperature, 3.00--4.68~dex in surface gravity and -1.43 -- 
+0.30~dex in global metallicity.

\begin{table}
\caption{The sample of the observed stars. The full table 
is available in the electronic version.}
\label{tab:samp}
\begin{tabular}{lcc}
\hline 
Name & V (mag) & Spectral type \\ 
\hline 
HD 236373 	            &  10.22	&	G0V \\
1RXS J003845.9+332534 	&  10.33	&	G2V \\
\rm{[BHG88]} 40 1943 	&  14.77	&	G2V \\
HD 4602 	                &   9.04	&	G0V \\
BD+59 140 	            &  10.37	&	G2V \\
TYC 4497-874-1        	&  10.92	&	G3V \\
HD 5649 	                &   8.70	&	G0V \\
TYC 4017-1351-1       	&  11.22	&	G0V \\
BD+60 167        	    &  10.34	&	G0V \\
BD+02 189 	            &  10.29	&	G1V \\
\hline
\end{tabular}
\end{table}

\begin{table*}
\caption{The atmospheric parameters for the 64 reference stars.} 
\label{tab:refs}
\begin{tabular}{lrrrlrrrlrrr}
\hline
Star  & $T_{\rm eff}$  & $\log{g}$  & [M/H] &  Star  & $T_{\rm eff}$  & $\log{g}$  & [M/H] &  Star  & $T_{\rm eff}$  & $\log{g}$  & [M/H]\\
& (K) & (dex) & (dex)& & (K) & (dex) & (dex) & & (K) & (dex) & (dex) \\\hline
HD3651    & 5205  &  4.49  &  0.10 & HD99285  & 6599 & 3.84 & -0.22 & HD134083  &  6537 &  4.31 &  0.02  \\
HD3765    & 5034  &  4.53  &  0.05 & HD100180 & 5927 & 4.25 & -0.06 & HD134113  &  5680 &  4.06 & -0.78  \\
HD15335   & 5858  &  3.93  & -0.18 & HD100563 & 6401 & 4.31 &  0.05 & HD136064  &  6121 &  4.03 & -0.03  \\
HD18757   & 5685  &  4.36  & -0.29 & HD101606 & 6134 & 3.98 & -0.75 & HD137052  &  6423 &  3.94 & -0.09  \\
HD18803   & 5658  &  4.46  &  0.13 & HD102574 & 6030 & 3.92 &  0.16 & HD139457  &  5954 &  4.05 & -0.51  \\
HD25621   & 6251  &  3.95  &  0.01 & HD106156 & 5464 & 4.68 &  0.18 & HD142357  &  6475 &  3.44 & -0.02  \\
HD28271   & 6160  &  3.85  & -0.10 & HD114606 & 5610 & 4.28 & -0.48 & HD142860  &  6286 &  4.14 & -0.20  \\
HD29645   & 5985  &  4.04  &  0.06 & HD117176 & 5527 & 3.95 & -0.06 & HD144284  &  6309 &  4.13 &  0.20  \\
HD33256   & 6242  &  3.99  & -0.36 & HD117361 & 6789 & 3.95 & -0.27 & HD145976  &  6720 &  4.10 &  0.01  \\
HD33608   & 6489  &  4.08  &  0.22 & HD120136 & 6445 & 4.30 &  0.27 & HD149414  &  5055 &  4.40 & -1.31  \\
HD35984   & 6175  &  3.68  & -0.07 & HD122742 & 5509 & 4.39 &  0.00 & HD149996  &  5662 &  4.10 & -0.53  \\
HD43386   & 6480  &  4.27  & -0.06 & HD125184 & 5659 & 4.11 &  0.27 & HD150012  &  6380 &  3.80 &  0.05  \\
HD61295   & 6987  &  3.05  &  0.25 & HD126512 & 5758 & 4.05 & -0.62 & HD150177  &  6096 &  3.95 & -0.58  \\
HD67228   & 5814  &  4.00  &  0.11 & HD126660 & 6322 & 4.27 & -0.04 & HD155646  &  6180 &  3.84 & -0.13  \\
HD76292   & 6866  &  3.77  & -0.22 & HD126681 & 5522 & 4.58 & -1.18 & HD157373  &  6427 &  4.08 & -0.43  \\
HD87646   & 5961  &  4.41  &  0.30 & HD127334 & 5651 & 4.15 &  0.18 & HD157856  &  6309 &  3.93 & -0.18  \\
HD87822   & 6597  &  4.10  &  0.17 & HD128167 & 6712 & 4.32 & -0.37 & HD159332  &  6184 &  3.85 & -0.23  \\
HD88986   & 5827  &  4.13  &  0.03 & HD128959 & 5478 & 3.00 & -0.92 & HD185144  &  5268 &  4.49 & -0.23  \\
HD91752   & 6423  &  4.03  & -0.25 & HD130087 & 6040 & 4.34 &  0.26 & HD190228  &  5306 &  3.83 & -0.26  \\
HD94028   & 5963  &  4.13  & -1.43 & HD130945 & 6431 & 4.06 &  0.06 & HD222404  &  4857 &  3.23 &  0.09  \\
HD95128   & 5861  &  4.30  &  0.00 & HD131156 & 5457 & 4.52 & -0.14 &           &       &       &        \\
HD99028   & 6739  &  3.98  &  0.06 & HD132375 & 6273 & 4.16 &  0.01 &           &       &       &        \\ 
\hline
\end{tabular}
\end{table*}

Since we know with great accuracy the photospheric parameters of the Sun, we also observed, in February 2013, its 
spectrum reflected by the asteroids Vesta and Ceres, using the same observational set-up and a 
CCD600 camera which provides a slightly larger dispersion of 1.0~\AA~pixel$^{-1}$. At the time of the observation, 
they had V=7.71 and 8.05~mag, respectively. We took five exposures of each target, for a total of 780 and 750 seconds, respectively, providing S/N$\sim$60.

\begin{figure*}
\includegraphics[width = 170mm]{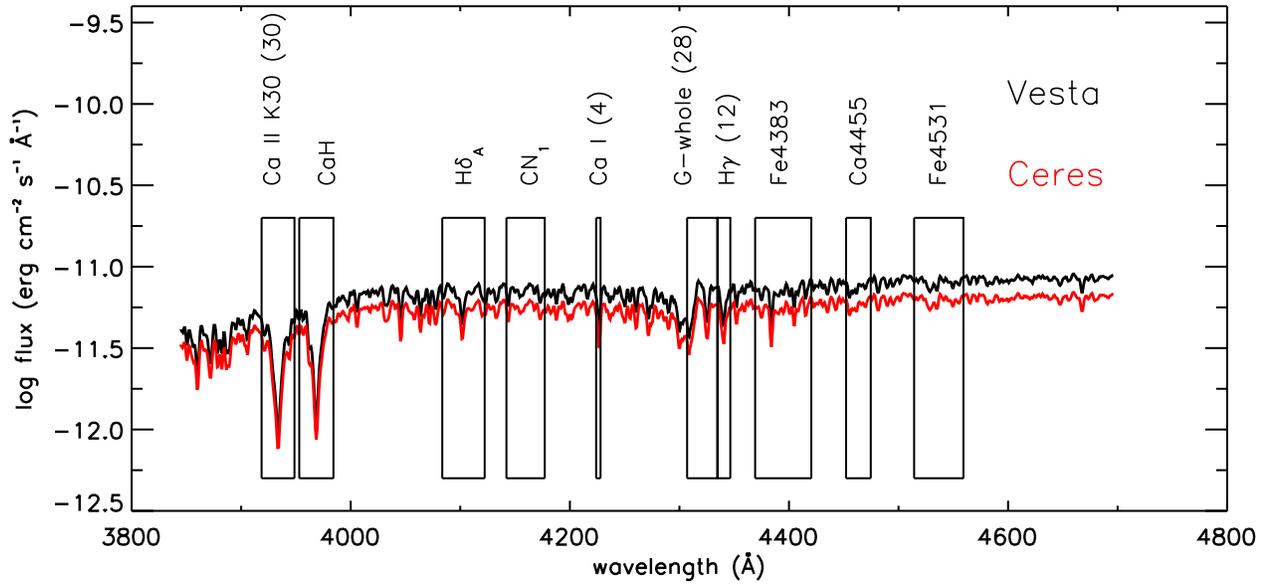}
\caption{Solar SED from the observations of the asteroids Vesta (black, upper spectrum) and Ceres (red, low spectrum).
The rectangles represent the central bands of the 10 
indices of Table~\ref{tab:ind}.}
\label{fig:sol}
\end{figure*}

\begin{figure*}
\includegraphics[width = 170mm]{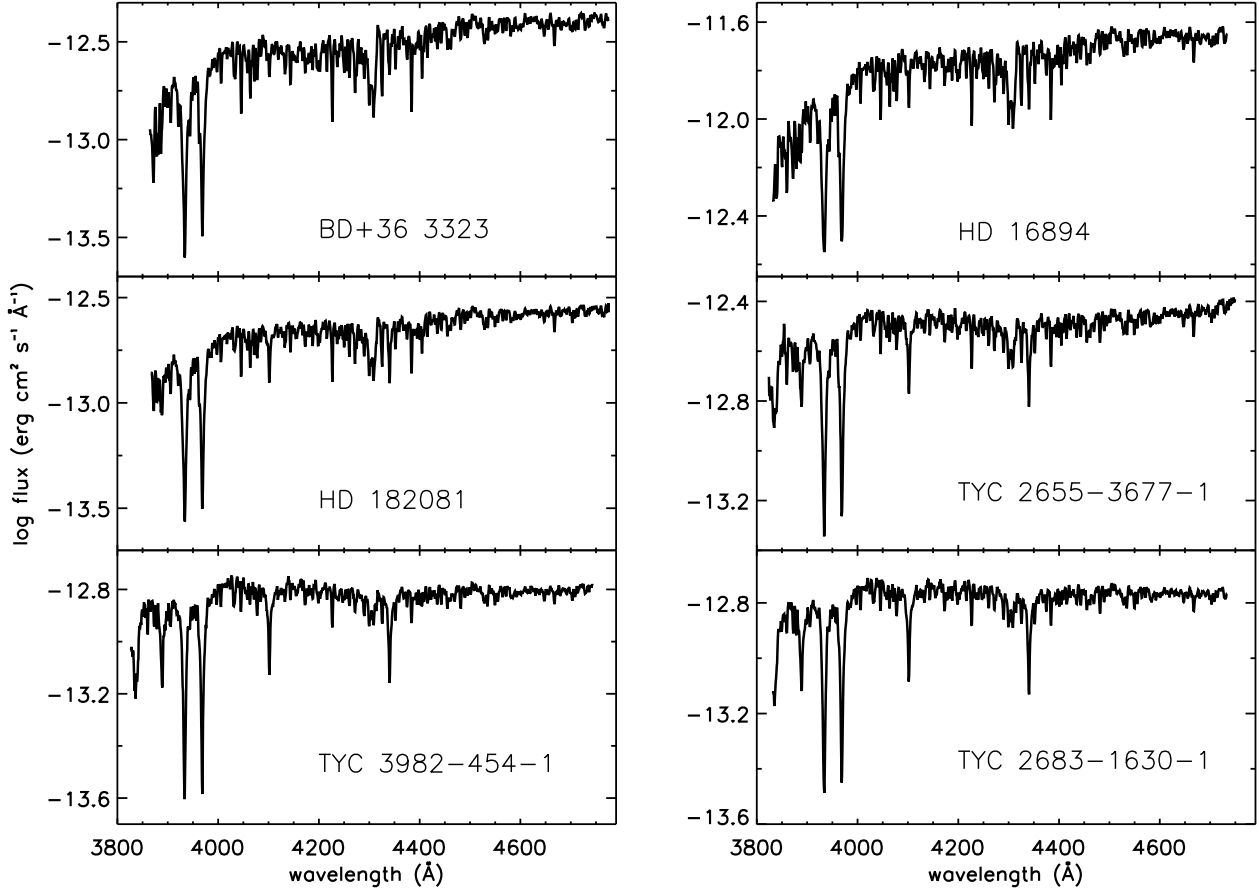}
\caption{SEDs of an illustrative sub-set of stars in our sample.}
\label{fig:seds}
\end{figure*}

The data were reduced following the standard procedure in IRAF: bias subtraction, flat field correction, cosmic ray removal, wavelength calibration (by means of an internal HeAr lamp), and flux calibration (using spectrophotometric standard stars from the ESO list. We then shifted all spectra to the rest frame, using
an average radial velocity obtained from measuring the shift of several absorption lines along each spectrum. Finally, for each star we co-added its multiple spectra, weighted by their mean S/N.
We show the spectral energy distributions (SEDs) of Vesta and Ceres in Fig~\ref{fig:sol}, while in Fig.~\ref{fig:seds} we present the spectra of some  representative objects.

\section{The library of synthetic stellar spectra}

With the goal of determining the atmospheric parameters of the target stars, we need a 
theoretical counterpart. We adopted the library of high spectral resolution SEDs of 
\citet{munari05}, which is based on the ATLAS model atmospheres of \citet{castelli03} and 
covers the 2500--10500~\AA\ wavelength range. \citet{bertone04} showed that ATLAS 
theoretical SEDs \citep{kurucz93,castelli03} are suitable to 
match the spectra of G-type stars, more so when the 
line blending at the spectral resolution of our observations dilutes the potential inconsintencies
that might be present in individual lines \citep{bertone08}. 
From the whole collection of \citet{munari05} SEDs, we extracted a subset of SEDs, hereafter called the Munari's library, using the following criteria: resolving power $R=20\,000$, $4750 \leq T_{\rm eff} \leq 7000$~K, $3.0 \leq \log{g} \leq 5.0$~dex, 
$-2.5 \leq \rm{[M/H]} \leq 0.5$~dex, solar-scaled abundances, microturbulence $\xi = 2$~km~s$^{-1}$, and rotational velocity $V_{\rm rot}= 0$~km~s$^{-1}$.
We then degraded all the SEDs to match the observational spectral resolution with a Gaussian convolution.

\section{The spectroscopic indices} 

In order to quantify the flux absorption of the most relevant spectral features, we considered the 41 spectral indices defined in the works of \citet{worthey97}, \citet{trager98}, \citet{carretero07}, and \citet{lee08} that fall within the spectral range of the 
observations. These indices are constructed with 3 wavelength bands: a central one, that includes
the selected spectral feature, and two side bands, conveniently placed where the line blanketing is minimum, that 
are used to define a pseudo-continuum level.
For the sake of homogeneity, we computed the values $I$ of all indices following the definition of 
the Lick/IDS system \citep{trager98}. 
The integrated fluxes within the side bands provide, 
through a linear fit, the pseudo-continuum level 
for the central band. The index is obtained from
\begin{equation}
I = \int_{\lambda_1}^{\lambda_2}\left(1 - \frac{F_{\rm cb}(\lambda)}{F_{c}(\lambda)}\right)d\lambda
\label{l2}\, ,
\end{equation}
where $F_{\rm cb}$ is the flux in the central band, whose wavelength limits are $\lambda_1$ and $\lambda_2$, and 
$F_{\rm c}$ is the flux of the pseudo-continuum at the central band. All results are therefore given as a pseudo-equivalent width in angstroms.

We computed the full set of indices for all the observed stars 
and the synthetic SEDs of the Munari's library. However, we made use of only 10 indices to determine the 
stellar parameters.  We selected these 10 indices as the most promising for the determination of atmospheric parameters, as described in Sect.~\ref{sec:selind}. We report the definition of the wavelength bands of these indices in Table~\ref{tab:ind}, along with their typical error (see Sect.~\ref{sec:selind}).

\begin{table*}
\begin{minipage}{126mm}
\caption{Wavelength band definition for 10 spectroscopic indices. Last column provides that reference where 
the index was defined: S = \citet{lee08}, L = \citet{trager98}, W = \citet{worthey97} y C = \citet{carretero07}.}
\label{tab:ind}
\begin{tabular}{lccccc}
\hline
Name  & Central Band     & Blue Band         & Red Band         & $\bar{\sigma_I}$ & Ref. \\
& $\lambda$ (\AA)           &$\lambda$ (\AA)    &$\lambda$ (\AA)     & (\AA)      &    \\\hline
Ca II K30 (30)     & 3918.600 - 3948.600 & 3907.500 - 3912.500 & 4007.500 - 4012.500 & 0.18 & S \\ 
CaH              & 3953.000 - 3984.200 & 3890.300 - 3913.100 & 4008.700 - 4029.200 & 0.14 & C \\
H$\delta_{A}$  & 4083.500 - 4122.250 & 4041.600 - 4079.750 & 4128.500 - 4161.000 & 0.15 & W \\
CN$_1$           & 4142.125 - 4177.125 & 4080.125 - 4117.625 & 4244.125 - 4284.125 & 0.15 & L \\
Ca I (4)              & 4224.000 - 4228.000 & 4208.000 - 4214.000 & 4230.000 - 4234.000 & 0.04 & S \\
G-whole (28)         & 4307.000 - 4335.000 & 4090.000 - 4102.000 & 4500.000 - 4514.000 & 0.13 & S \\
H$\gamma$ {(12)}   & 4334.500 - 4346.500 & 4247.000 - 4267.000 & 4415.000 - 4435.000 & 0.07 & S \\
Fe4383           & 4369.125 - 4420.375 & 4359.125 - 4370.375 & 4442.875 - 4455.375 & 0.23 & L \\
Ca4455           & 4452.125 - 4474.625 & 4445.875 - 4454.625 & 4477.125 - 4492.125 & 0.11 & L \\
Fe4531           & 4514.250 - 4559.250 & 4504.250 - 4514.250 & 4560.500 - 4579.250 & 0.18 & L \\
\hline
\end{tabular}
\end{minipage}
\end{table*}

\subsection{Transformation of the indices to the observational system}
\label{sec:calib}
Theoretical spectra do not perfectly reproduce the observations \citep[e.g.][]{bertone04, bertone08, coelho14}, as they are 
affected by systematic effects (physical and geometrical approximations, inaccurate atomic parameters for opacity 
computation, etc.). Therefore, we must first transform the theoretical indices from the Munari's library to the 
observational system by using the set of reference stars. 
For each of the these objects, we made a trilinear interpolation in the ($T_{\rm eff}$, $\log{g}$, [M/H]) parameter 
space to produce its synthetic spectrum. Then, for each index, we carried out a 
least square  linear fitting of the synthetic vs. observed indices  of the reference stars to transform 
the theoretical indices as:
\begin{equation}
I_{\rm {teo,cal}} = \frac{I_{\rm teo}-b}{m} \, ,
\end{equation}
where $I_{\rm {teo,cal}}$ is the calibrated theoretical index, I$_{\rm teo}$ is the original theoretical index, and 
$b$ 
and $m$ are the y-intercept and slope of the linear fitting. To exclude unreliable outliers, we adopted an iterative 
3$\sigma$ clipping method.

\subsection{Selection and properties of the set of suitable indices} 
\label{sec:selind}
In order to obtain the most accurate and precise photospheric parameters, we carried out a meticulous index selection to 
extract, from the pool of 41 indices, those that generate a good degree of ``orthogonality'' in the 
($T_{\rm eff}$, $\log{g}$, [M/H]) parameter space and that are relatively well reproduced by the uncalibrated 
theoretical values. We realized that the inclusion of inadequate indices affects both accuracy and precision of the 
results. 

The first step in the process was to select only those indices with slopes $0.6<m<1.4$ of the calibration equation, 
so to immediately discard the less reliable indices. Several indices measure the same spectral feature, but 
with different definitions of the three wavelength bands. In those cases, we picked up just one index, based on the 
visual inspection of the behaviour of the indices against variations in $T_{\rm eff}$, $\log{g}$ and [M/H]: for 
instance, in the case of the Balmer H$\delta$ line, we chose the index that maximized the sensitivity to 
temperature, while minimized the dependence on surface gravity and metallicity.
Finally, we used the observed solar spectra that we collected from Vesta and Ceres as a test bench. We considered 
many combinations of the sub-set of indices selected at this point and we performed a chi-square analysis to 
determine the photospheric parameters of the Sun, by comparing the observed and calibrated theoretical indices, and 
we chose the combination that provided the best match with the accepted parameters of the Sun ($T_{\rm eff}$, 
$\log{g}$, [M/H])=(5777~K, 4.44, 0.0). The best result was given by the set of the following 10 indices: Ca 
II K30 {(30)}, CaH, H$\delta_{\rm A}$, CN$_1$, Ca I (4), G-whole (28), H$\gamma$ (12), Fe4383, Ca4455, and Fe4531; 
they yielded, for both spectra, ($T_{\rm eff}$, $\log{g}$, [M/H])= (5750,4.50,-0.02).

\begin{figure}
\includegraphics[width = 84mm]{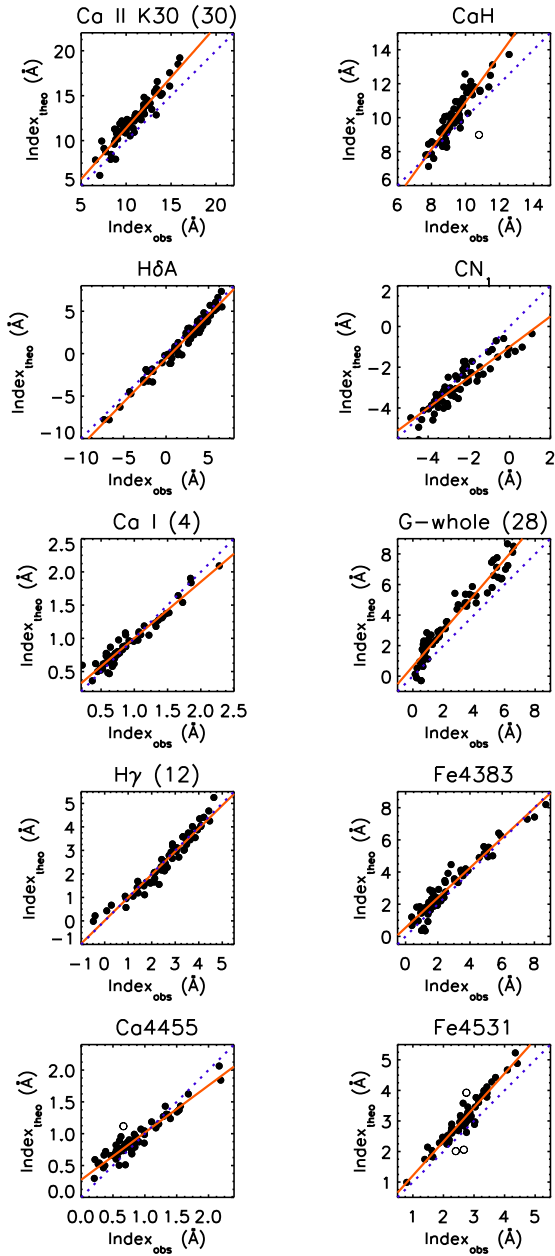}
\caption{Theoretical vs. observed indices for the set of {\it reference stars} for the selected set of indices. The 
dotted line indicates the bisector of the plane, while the solid line shows the best linear fit of the data. The open dots are the data rejected by 3$\sigma$ clipping.}
\label{fig:calib}
\end{figure}

For this set of indices, we present in Fig.~\ref{fig:calib} the plots of the theoretical vs. observed indices for the reference stars, along with the linear best 
fit, from which we derived the transformation equation, whose $b$ and $m$ parameters are given in 
Table~\ref{tab:rcalib}. 
The two Balmer line indices and  Fe4383 are already very well reproduced by the synthetic Munari's spectra. The other indices need a larger correction to match the observations. In several cases, the linear fit crosses the bisector: this  indicates that there exists a combination of photospheric parameters for which the theoretical index matches the observed one. However, these combinations strongly vary from index to index, revealing that different spectral regions (or element opacity) are better reproduced at different, for instance, $T_{\rm eff}$.

\begin{table}
\caption{The slope, y-intercept, standard error between the fit 
and the observational data, and the number of stars rejected in the 3-$\sigma$ clipping method are of the 
best linear fit.}
\label{tab:rcalib}
\begin{tabular}{lcccc}
\hline 
Index           &       $m$ & $b$& $\sigma$ & r$_{3\sigma}$ \\ \hline
Ca II K30 (30)  &       1.134  &  0.043 &   0.857  &   0\\ 
CaH             &       1.387  & -2.963 &   0.578  &   1  \\
H$\delta_A$     &       1.026  & -0.599 &   0.627  &   0   \\   
CN$_{1}$         &      0.749  & -1.001 &   0.465  &   0  \\   
Ca I (4)        &       0.848  &  0.154 &   0.090  &   0  \\   
G-whole (28)    &       1.164  &  0.646 &   0.630  &   0  \\   
H$\gamma$ (12)  &       0.980  &  0.013 &   0.284  &   0  \\   
Fe4383          &       0.930  &  0.543 &   0.484  &   0  \\   
Ca4455          &       0.741  &  0.276 &   0.099  &   1  \\   
Fe4531          &       1.115  &  0.097 &   0.206  &   3  \\   
\hline
\end{tabular}
\end{table}

\begin{figure}
\includegraphics[width=84mm]{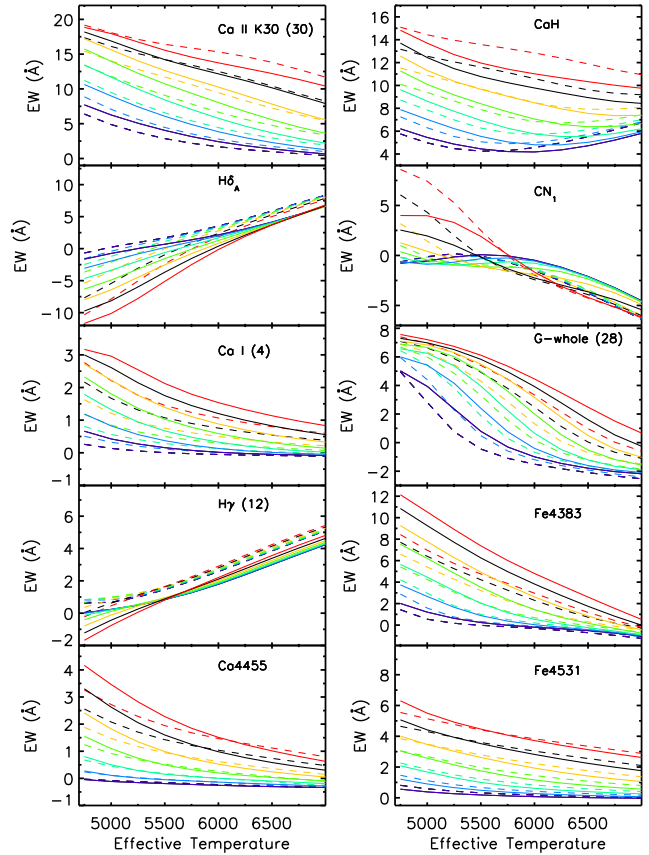}
\caption{Indices values as a function of stellar parameters. 
In each panel the surface gravity is fixed at the values
$\log g$~=~3.0 (dashed lines) and 5.0~dex (solid lines).
Different colours represent different metallicities, from 
\rm{[M/H]}~=~-2.5~dex (purple) 
to 0.5~dex (red), with a step of 0.5~dex.}
\label{f3}
\end{figure}

The transformed theoretical indices as a function of the three atmospheric parameters are depicted in Fig.~\ref{f3}, however, in order to understand how significant is the sensitivity of each index versus these parameters, we can 
compare its dynamical range with a typical error. We estimated this error with a Montecarlo method: we assumed a 
S/N = 50 at 4660~\AA, where the line blanketing is minimum, to simulate a photometric error for each synthetic 
spectrum of the Munari's library. For each spectrum, we run a set of 1000 realizations, where we added a 
randomly generated Gaussian photometric error and we computed the indices. The standard deviation of each index 
distribution provided the index error $\sigma_{I_{ijk}}$ for each combination of parameters. In Table~\ref{tab:ind}, 
we report the average error value $\bar{\sigma_I}$ over the whole Munari's library.

A quantification of the sensitivity of an index with respect to a given atmospheric parameter is provided by the 
ratio between the index range in the Munari's grid and the mean of $\sigma_{I_{ijk}}$, called {\em throw} 
\citep{worthey94}, at each mesh point of the plane formed by the other two parameters. 

\begin{figure*}
\centering
\begin{tabular}{cc}
\includegraphics[width = 78mm]{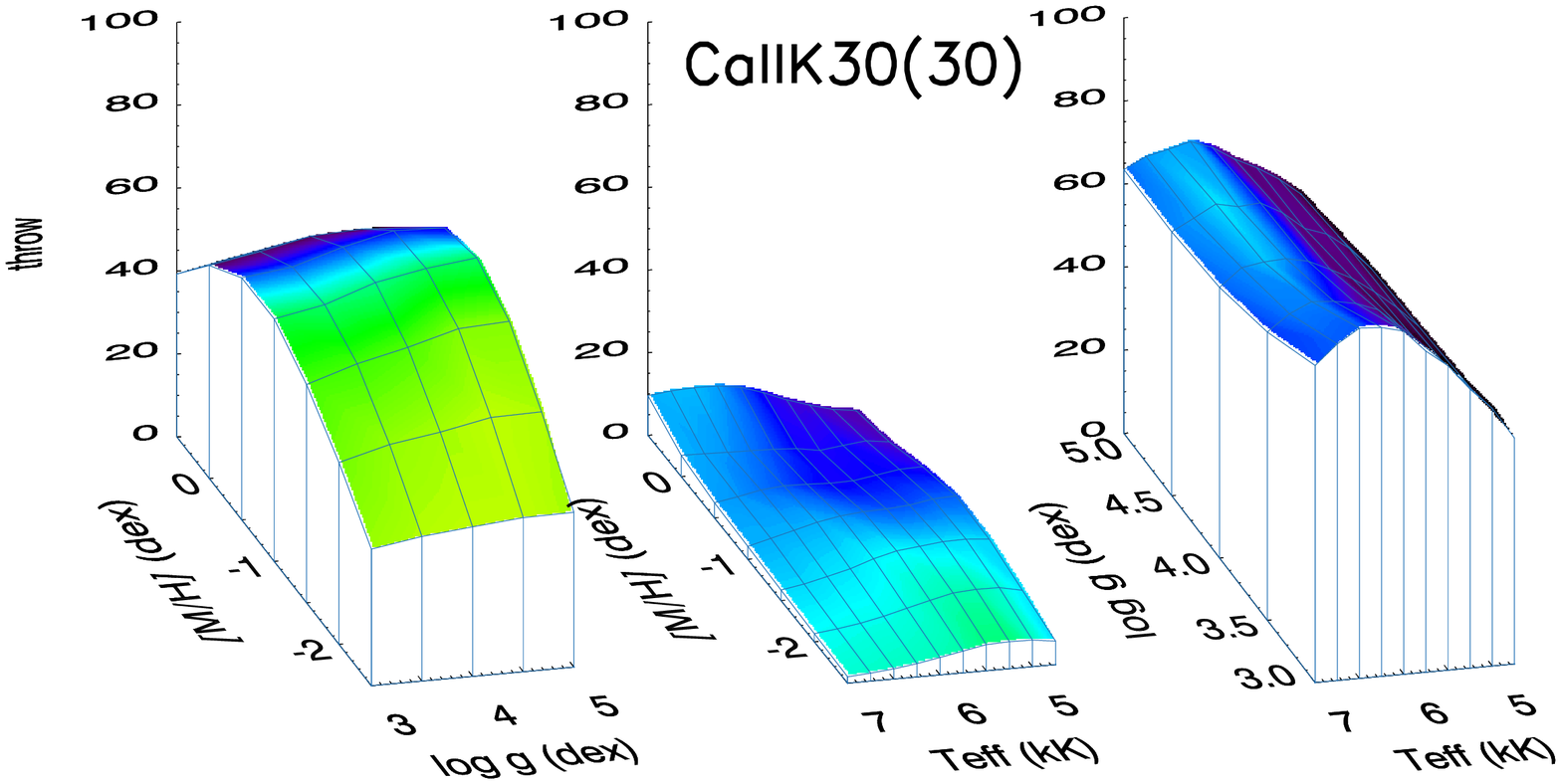} &
\includegraphics[width = 78mm]{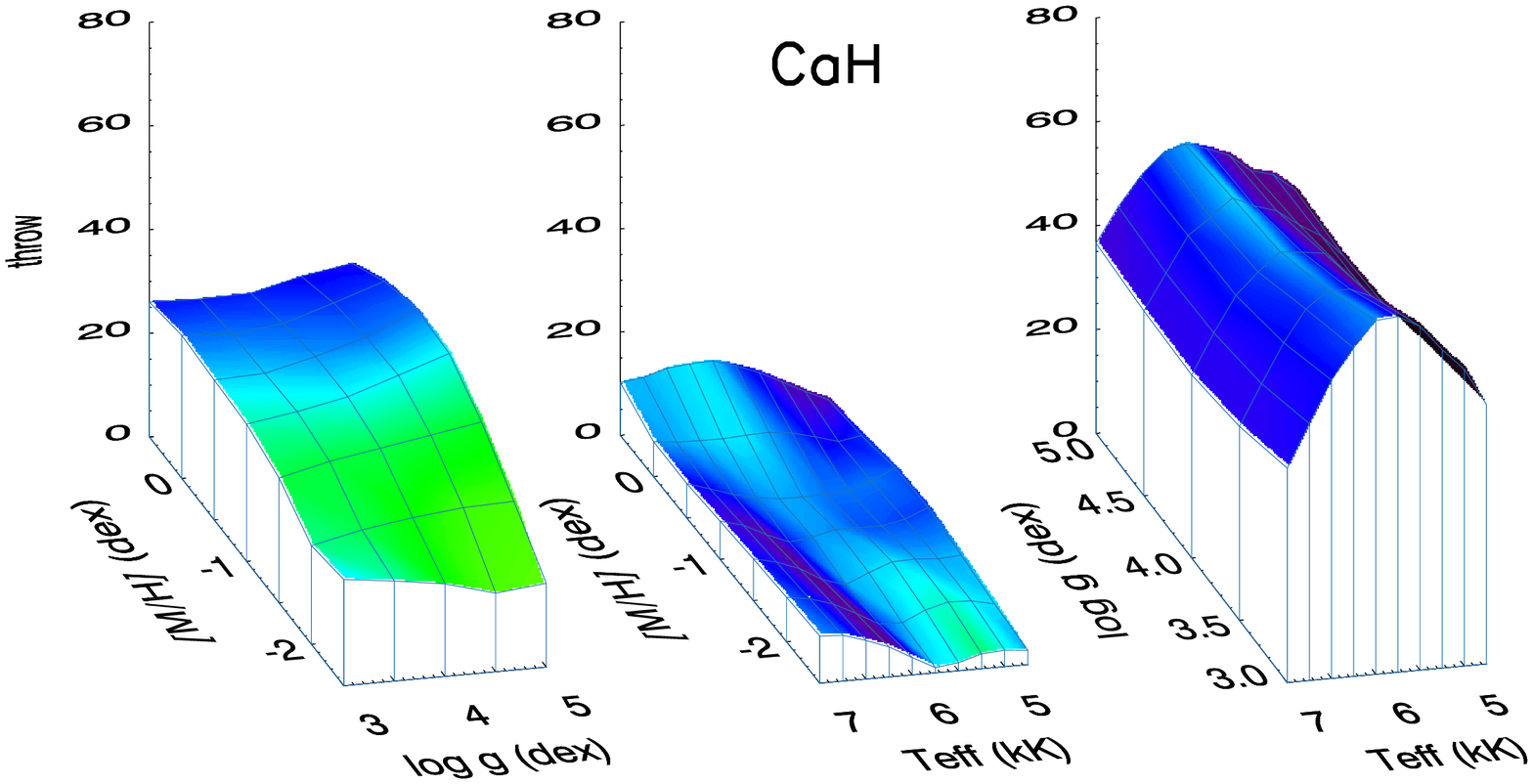} \\
\includegraphics[width = 78mm]{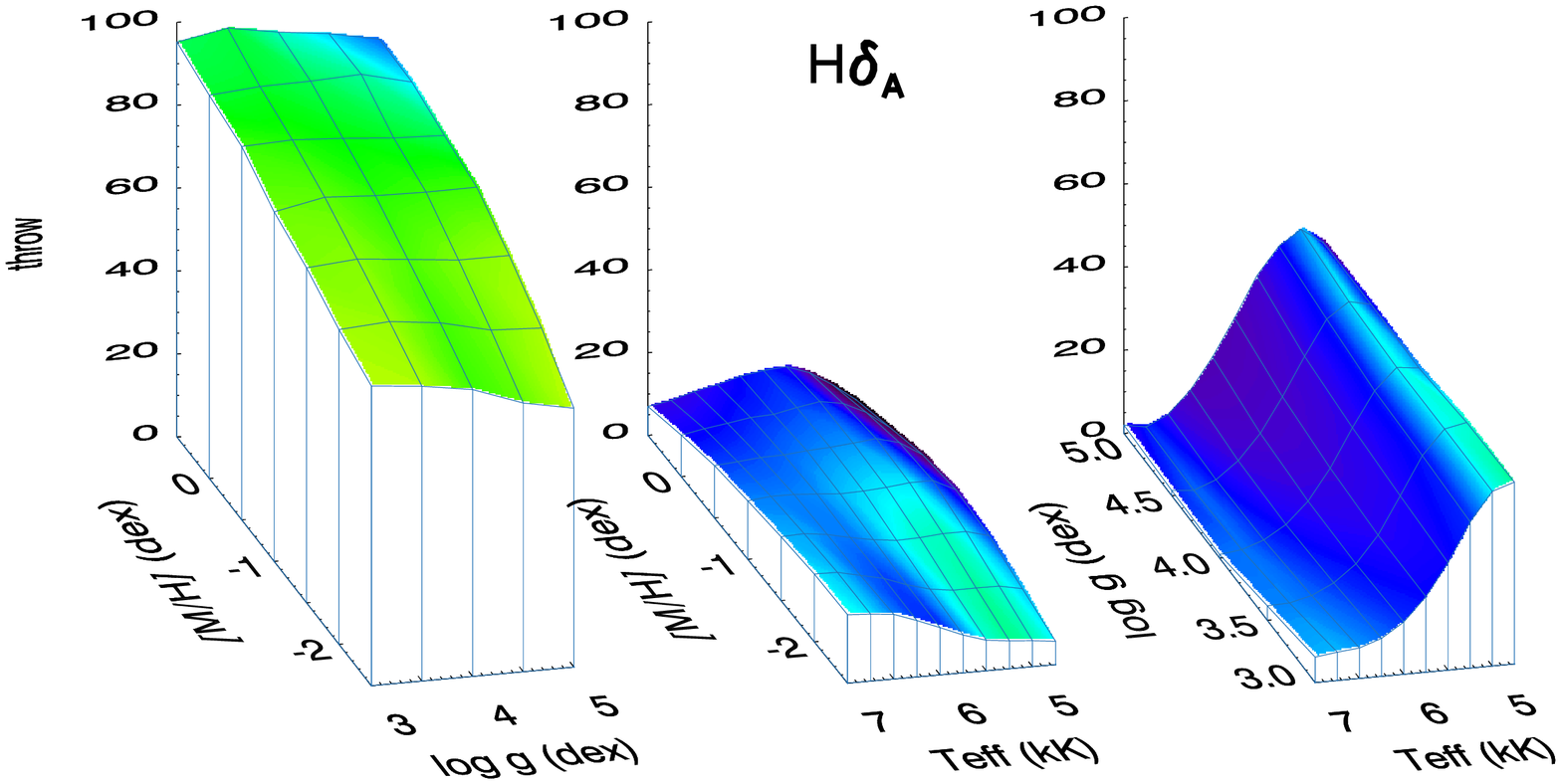} &
\includegraphics[width = 78mm]{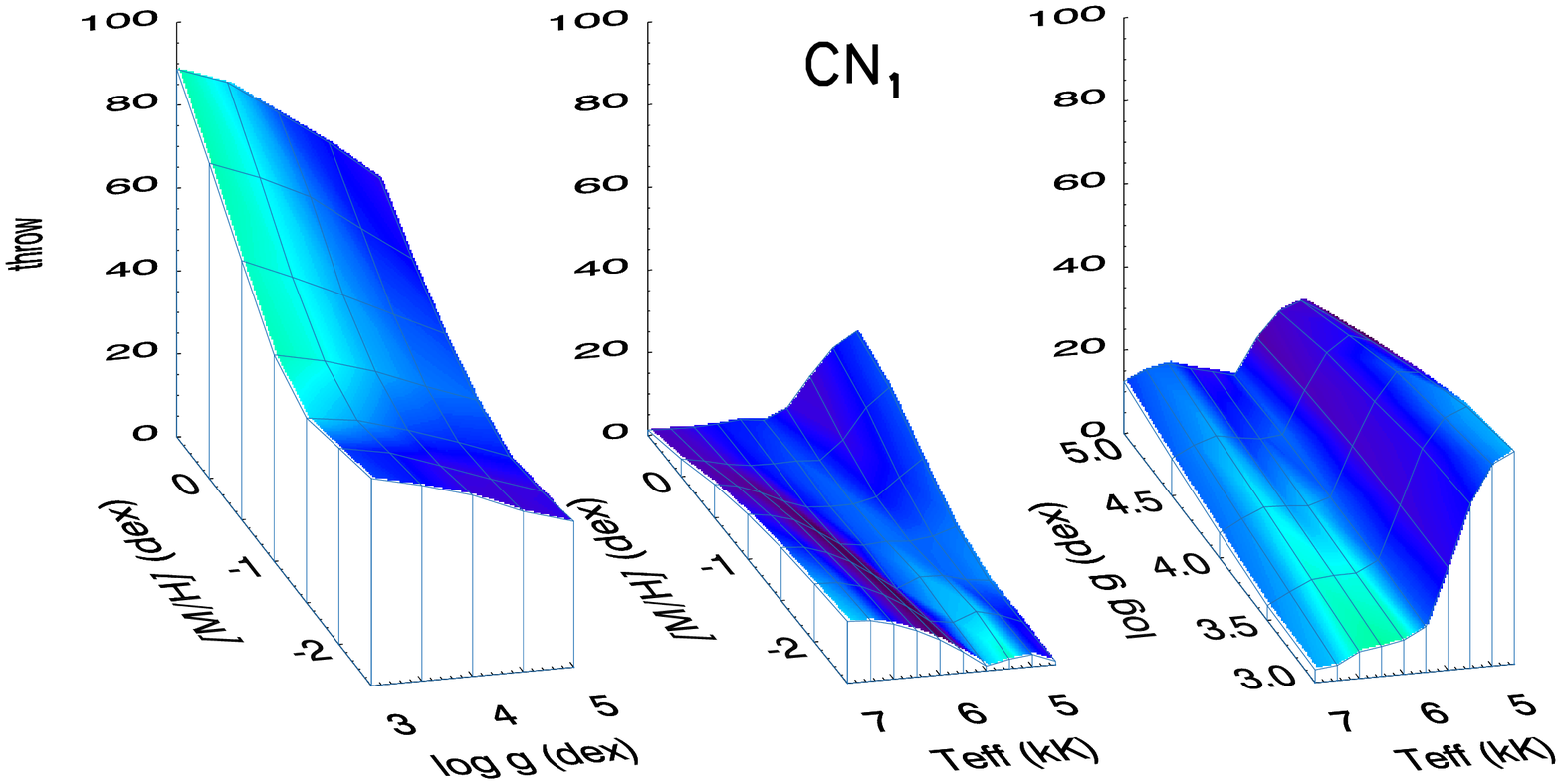} \\
\includegraphics[width = 78mm]{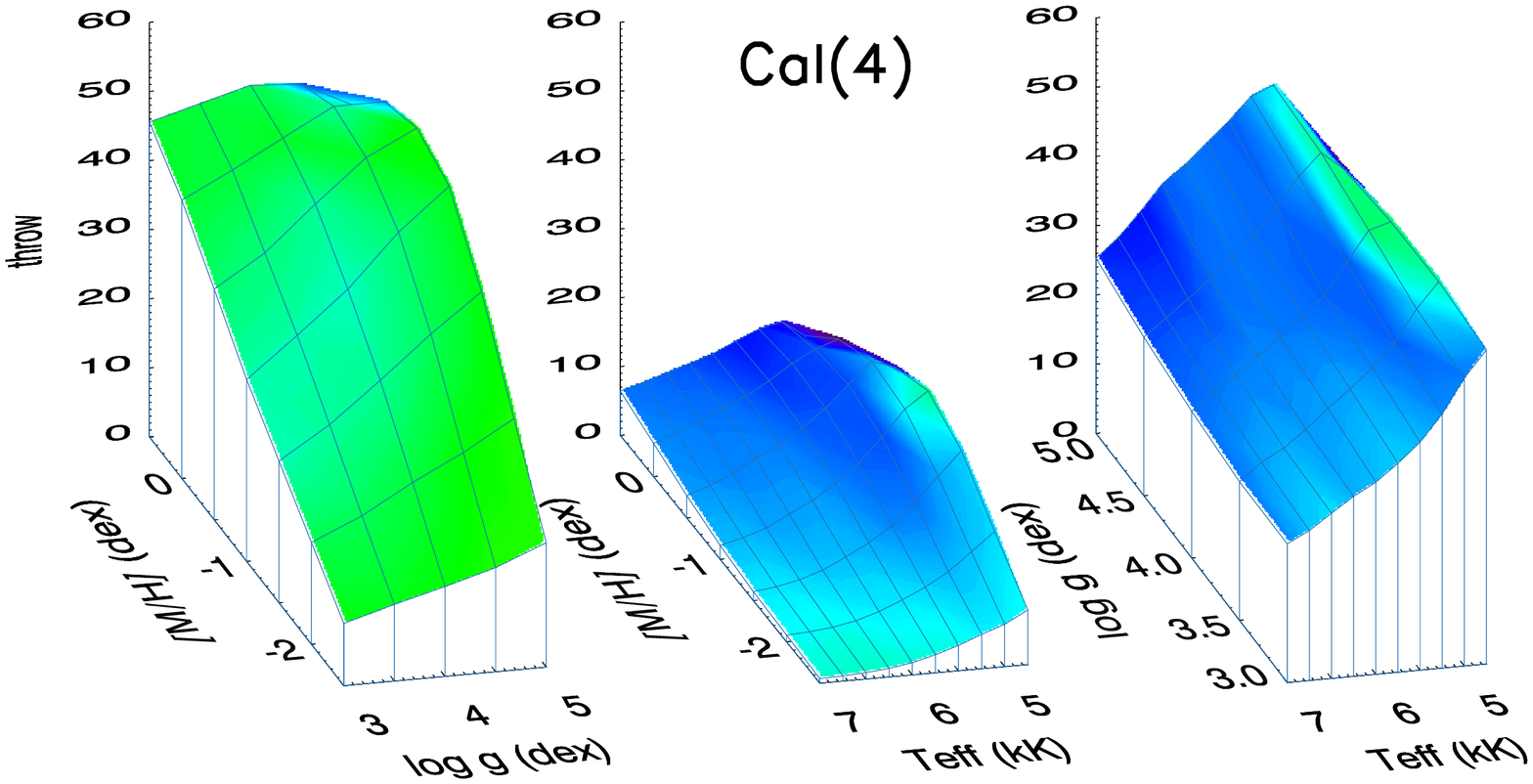} &
\includegraphics[width = 78mm]{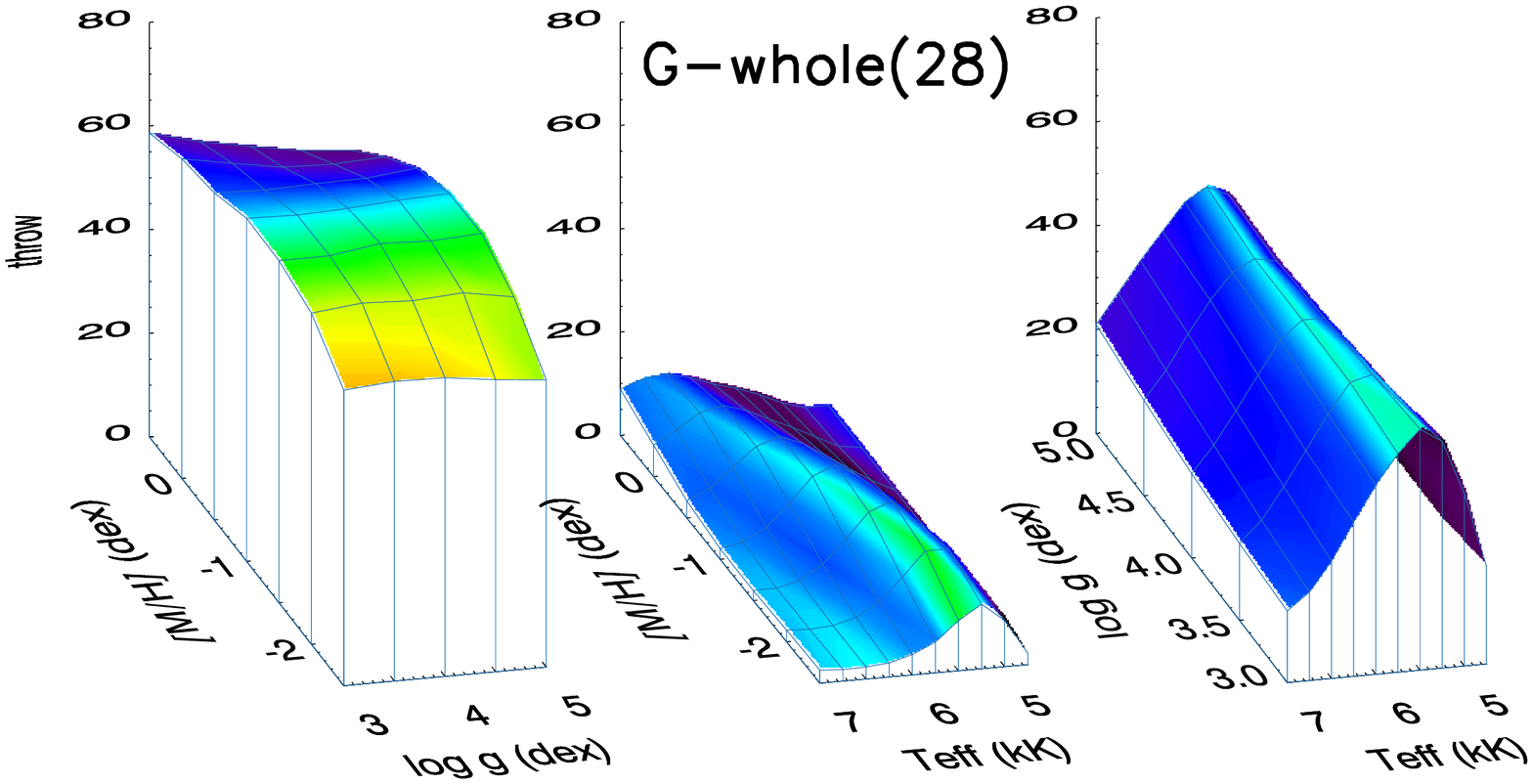} \\
\includegraphics[width = 78mm]{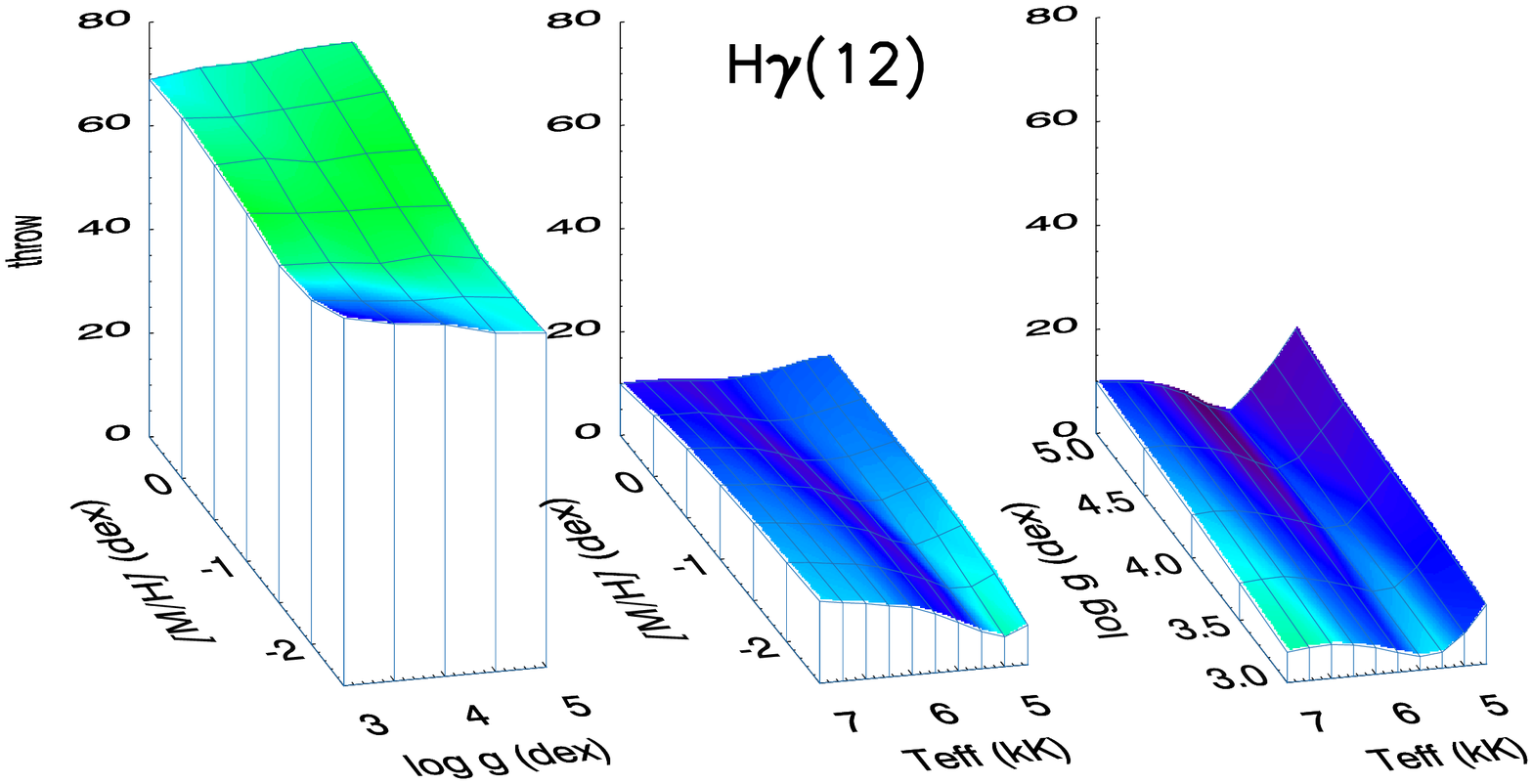} &
\includegraphics[width = 78mm]{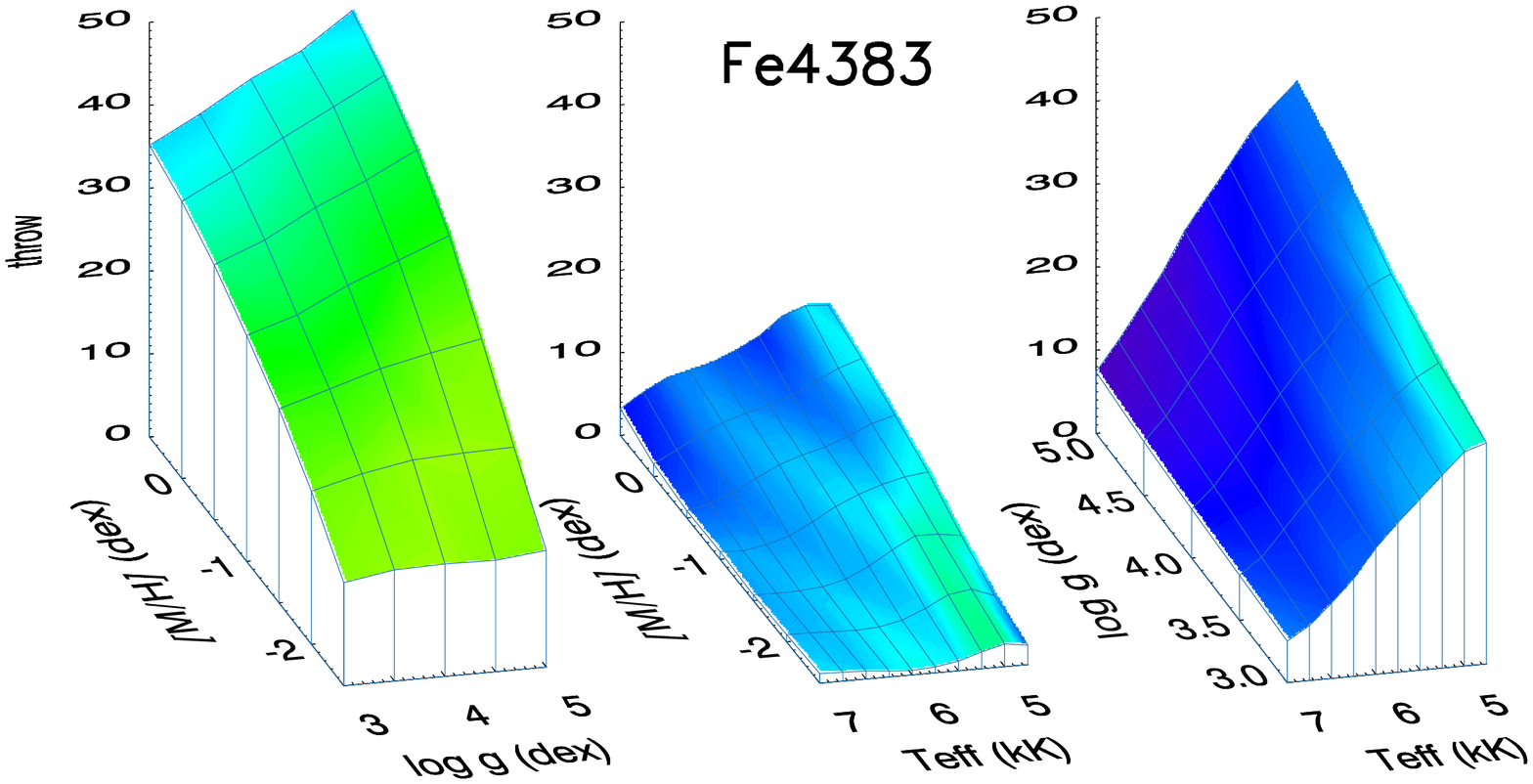} \\
\includegraphics[width = 78mm]{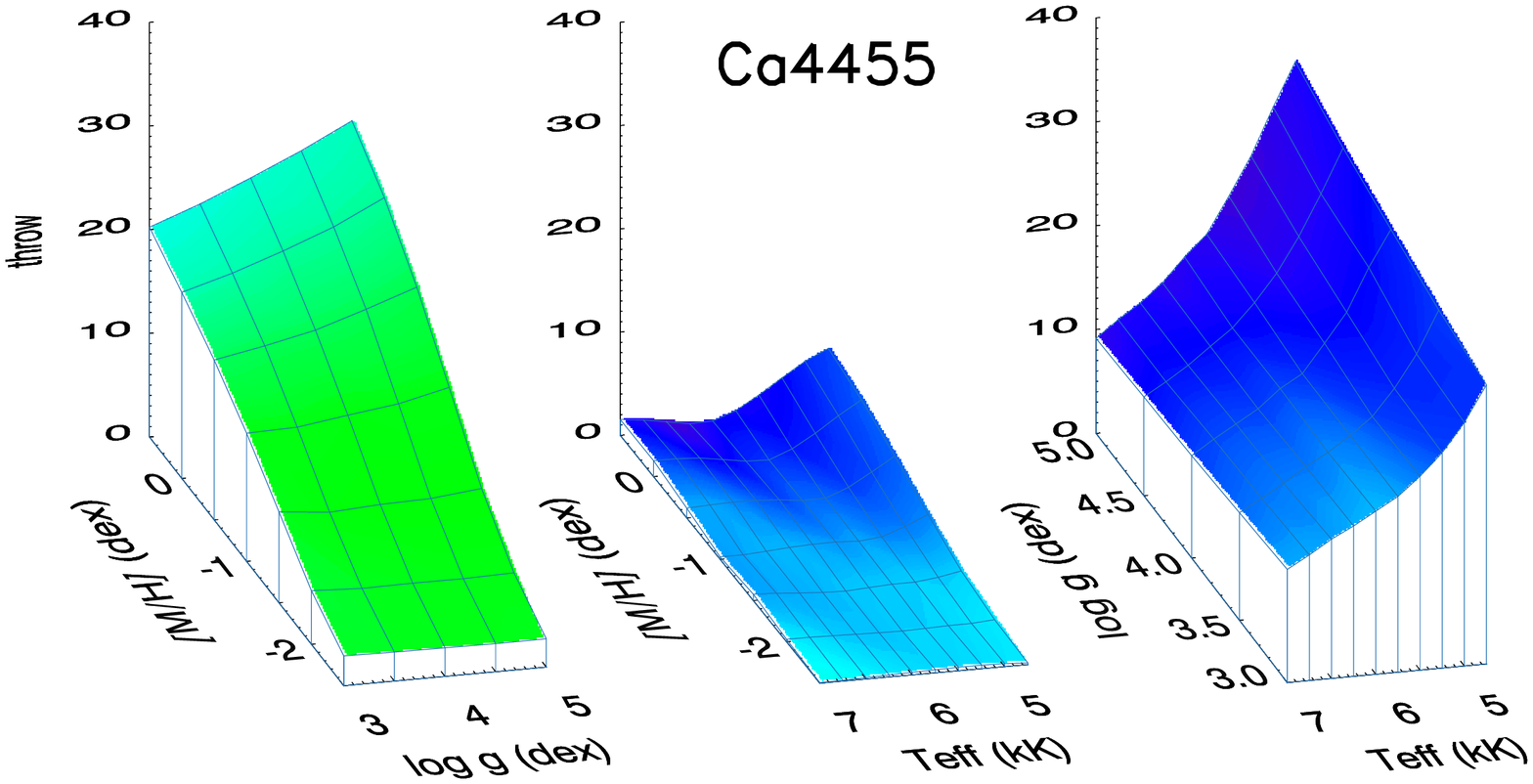} &
\includegraphics[width = 78mm]{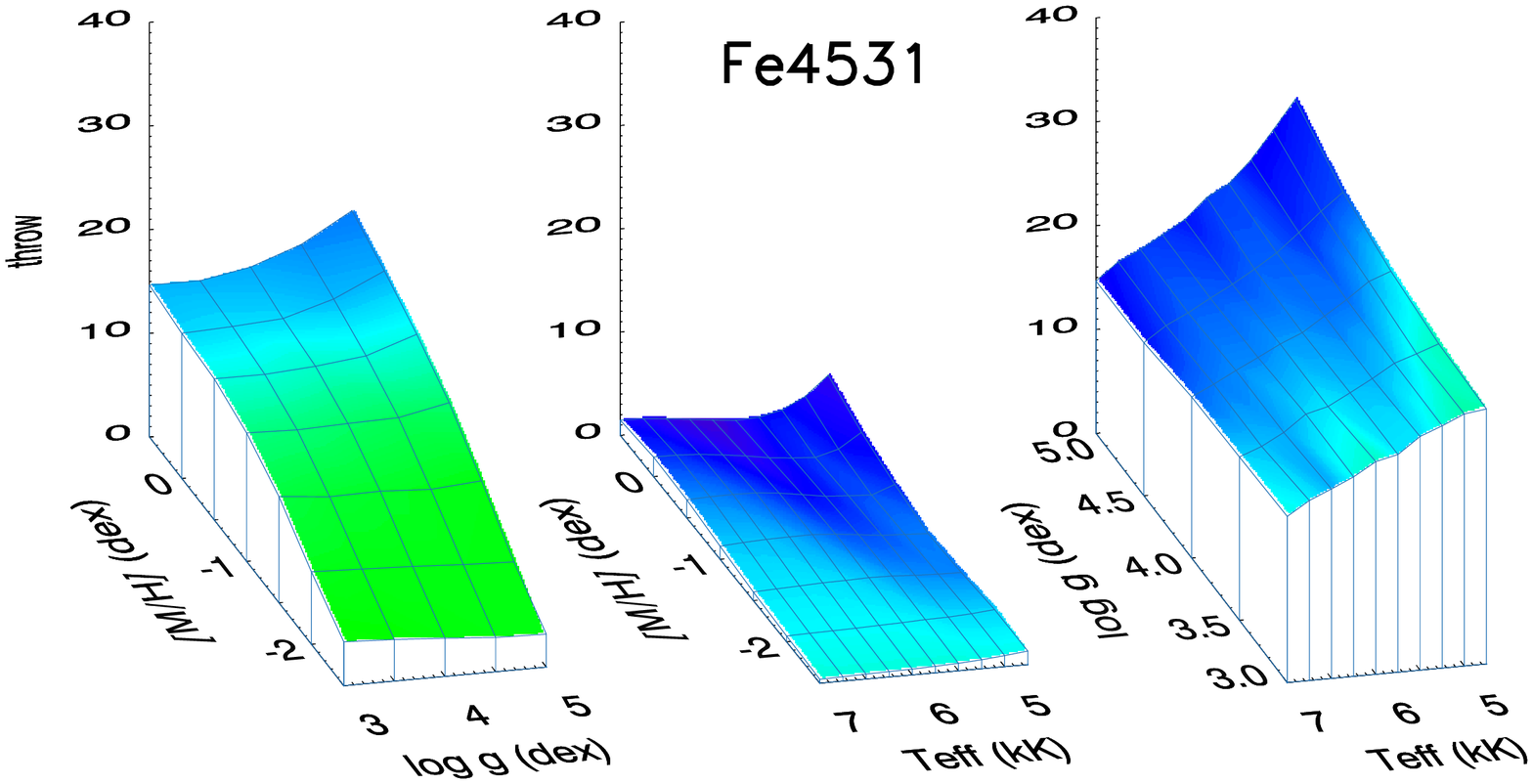} \\
\end{tabular}
\caption{Throw vs. $T_{\rm eff}$ (left plot of each index), $\log{g}$ (middle plot), and [M/H] (right plot). Note 
that the temperature scale is increasing from right to left.}
\label{fig:throw}
\end{figure*}

We present the {\em throw} of the 10 indices with respect to $T_{\rm eff}$, $\log{g}$, and [M/H] in 
Fig.~\ref{fig:throw}.
We observe a  variety of different behaviours: this diversity makes this combination of indices appropriate for a 
precise determination of the atmospheric parameters (notice, for instance, the complementarity between CN$_1$ and 
G-whole (28)). The sensitivity to $\log{g}$ is in general much lower than with respect to $T_{\rm eff}$ or
[M/H]. This behaviour will reflect into a higher error on the $\log{g}$ determination. However, the 
insensitivity to surface gravity makes some indices, as Fe4531 and Ca4455, better tools for determining the other two 
parameters.
The highest {\em throw} is reached by H$\delta_{\rm A}$, CN$_1$, and Ca II K30 ${(30)}$. A peculiar property of 
the latter index is that it shows the higher sensitivity to metallicity at higher temperatures. Conversely, the CaH and G-whole (28) indices show a peak around solar $T_{\rm eff}$, while for the other indices, the sensitivity to metallicity reaches 
the maximum at the lower temperature edge, where the overall opacity of metal lines is generally higher.
H$\gamma$ (12) should be a very effective temperature probe, since the {\em throw} is quite constant over the 
whole $\log{g}$--[M/H] plane and its dependence to  $\log{g}$ and $T_{\rm eff}$ is much lower; however, the two 
Balmer line indices have, on average, the highest $\log{g}$ {\em throws}. Furthermore, the presence of many metal 
lines inside its index bands makes H$\delta_{\rm A}$ strongly sensitive to metallicity for late-G stars. 

\begin{figure}
\centering
\includegraphics[width=\columnwidth]{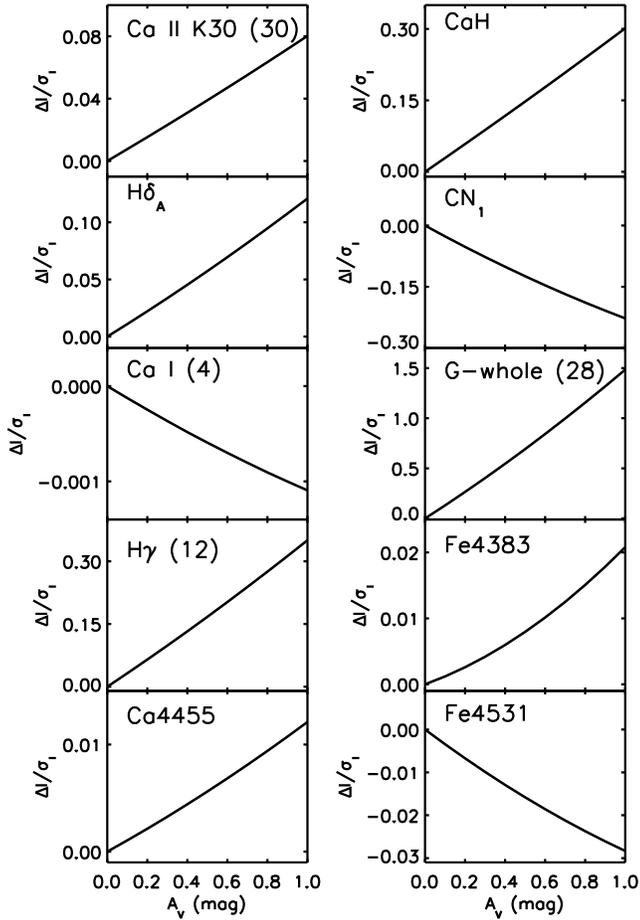}
\caption{Spectroscopic indices as a function of interstellar extinction. The value of each index is compared to the respective value of the index without extinction.}  
\label{fig:ext}
\end{figure}

\begin{figure}
\centering
\includegraphics[width=\columnwidth]{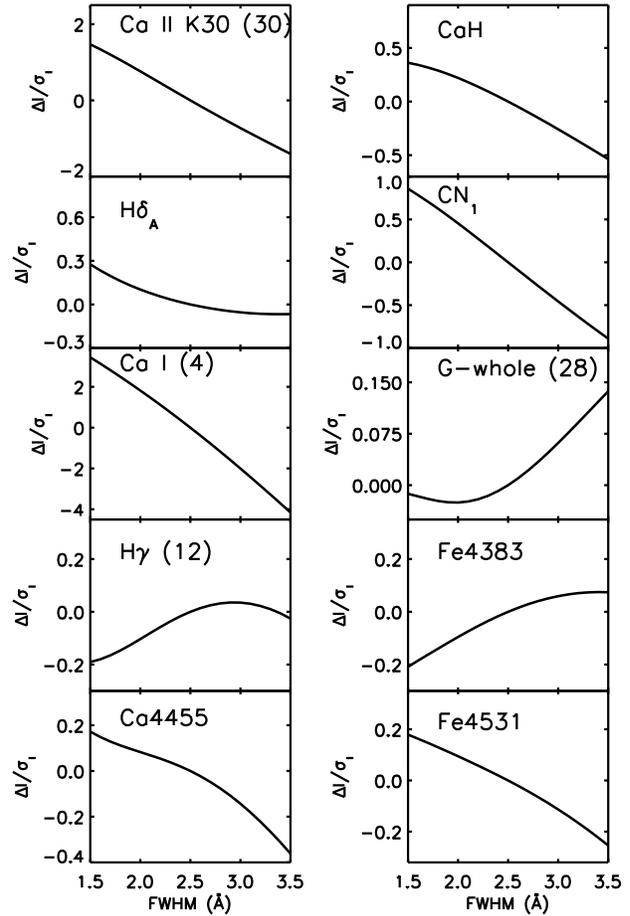}
\caption{Dependence of the indices to instrumental spectral resolution. The values of each index are compare to the 
respective value at 2.5~\AA\ of resolution.}  
\label{fig:resol}
\end{figure}

We also explored the dependence of the indices to extinction and instrumental spectral resolution, using as a 
reference the Munari's synthetic spectrum (5750,4.5,0.0). The effect of reddening is obtained by applying the 
curve of \cite{cardelli89}, increasing $A_V$  from 0 to 1~mag, and it is shown in Fig.~\ref{fig:ext}.
The only index that shows a variation of the same order of its typical error is G-whole(28), because its bands are defined over 
a wide wavelength interval (424~\AA). In terms of the {\em throw}, this means that an error of 1~mag in $A_V$ 
estimation for a star introduces an error of $\pm$1.5~{\em throws}, which has to be compared with the values depicted 
in Fig.~\ref{fig:throw} to translate this uncertainty in terms of atmospheric parameters error.
The effect of the extinction is much less for the other indices, being negligible for those defined in narrow 
wavelength intervals, such as CaI(4) or Ca4455.

Conversely, narrow band indices not affected by interstellar extinction might be subject to resolution effects.
We have explored the effects of instrumental resolution by degrading the above model flux from 1.5 to 
3.5~\AA\ FWHM at a step of 0.1 \AA~(see Fig.~\ref{fig:resol}). The most sensitive index is CaI(4), 
as its bandwidths are very narrow (4~\AA). Similarly, the narrow side bands (5~\AA\ width) are also the cause 
of the large variation shown by Ca II K30(30). However, the
bandwidth is not the only quantity that matters, as the other index that show a change larger than its {\em throw} 
is CN$_1$, whose bands are at least 30~\AA\ wide. In this case, the spectral morphology is also relevant since two 
strong feature dominated by Fe~\textsc{i} are placed very near the limits of the central band. This property makes this 
index also prone to be sensitive to the precision of the wavelength calibration of observed spectra.

\section{Determination of effective temperature, surface gravity, and metallicity.}
\label{sec:det}

Prior to the ultimate goal, we opted to extend the grid of synthetic indices. 
The parameter space coverage of Munari's library is too coarse, with steps of 250~K 
in  $T_{\rm eff}$, 0.5~dex in $\log{g}$ 
and 0.5~dex in [M/H]. We therefore performed a trilinear interpolation of the grid of 
calibrated indices to significantly reduce the steps to values that are smaller than 
the expected errors:
5~K in $T_{\rm eff}$, 0.05~dex in surface gravity and 0.02~dex in metallicity. This 
denser grid of indices includes more than 1.2 million entries.

We adopted a least squares method to determine the atmospheric parameters of the 
sample of target stars, by minimizing the statistic:
\begin{equation}
\label{eq:chi2}
\rm {\chi^2}=\frac{1}{n}\displaystyle \sum^{n}_{i=1} \frac{\left(I_{i,teo}-
I_{i,obs}\right)^2}{\sigma^{2}_{i,obs}}
\end{equation}
where $I_{i,teo}$ and $I_{i,obs}$ are the theoretical and the observational indices, 
respectively, $n$ 
is the total number of indices, and $\sigma_{i,obs}$ is the error associated to each 
index of the observed spectrum.
The combination of theoretical stellar parameters, whose indices provide the 
minimum $\chi^2$, was assigned to the corresponding star.

\begin{figure}
\includegraphics[width=80mm]{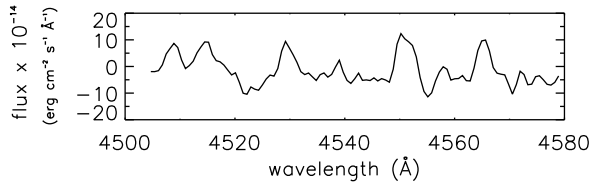}
\caption{Residual spectrum in the Fe4531 interval of the star BD+60~600, after the first iteration of the procedure described in Sect.~\ref{sec:det}}
\label{fig:resid}
\end{figure}

A critical point in this procedure is a correct estimation of the errors $\sigma^{2}_{i,obs}$ 
in Eq.~\ref{eq:chi2}: since 
the blue interval of G-type stars is populated with 
thousands of non-negligible absorption lines, there is no wavelength interval, at a resolution of FWHM=2.5~\AA, where the noise can be safely measured on the observed spectra. Therefore, we devised an 
iterative method that includes the following steps: 

\begin{table*}
\begin{minipage}{126mm}
\caption{Stellar atmospheric parameters. (The full version of the table is available in 
electronic form as Supporting Information in the online version of the article.)}
\label{tab:param}
\begin{tabular}{lccccccccc}
\hline
{Object} & {T$_{\tiny \mbox{eff}}$ (K) }& \multicolumn{2}{c}{$\sigma_{\tiny \mbox{T}_{\mbox{eff}}}$ (K)} & 
$\log g$ (dex)& \multicolumn{2}{c}{$\sigma_{\tiny \mbox{log g}}$ (dex)} & {$\left[\mbox{M/H}\right]$ (dex)} & \multicolumn{2}{c}{$\sigma_{\tiny \mbox{[M/H]}}$ (dex)} \\
   & & + & - & & + &- &  & + &- \\ 
\hline
Vesta        &   5750 & 35 & 50 & 4.50 & 0.15 & 0.25 & -0.02 & 0.04 & 0.04  \\
Ceres        &   5750 & 25 & 55 & 4.50 & 0.10 & 0.25 & -0.02 & 0.04 & 0.04  \\
HD 236373    &   6010 & 15 & 25 & 4.50 & 0.05 & 0.15 & -0.18 & 0.02 & 0.02  \\
1RXS J003845.9+332534  &   5815 & 50 & 50 & 4.70 & 0.25 & 0.20 & -0.26 & 0.06 & 0.06  \\
\rm{[BHG88]} 40 1943 &   5805 & 95 &140 & 3.85 & 0.50 & 0.70 & -0.30 & 0.14 & 0.16  \\  
\hline
\end{tabular}
\end{minipage}
\end{table*}

\begin{figure*}
\includegraphics[width=126mm]{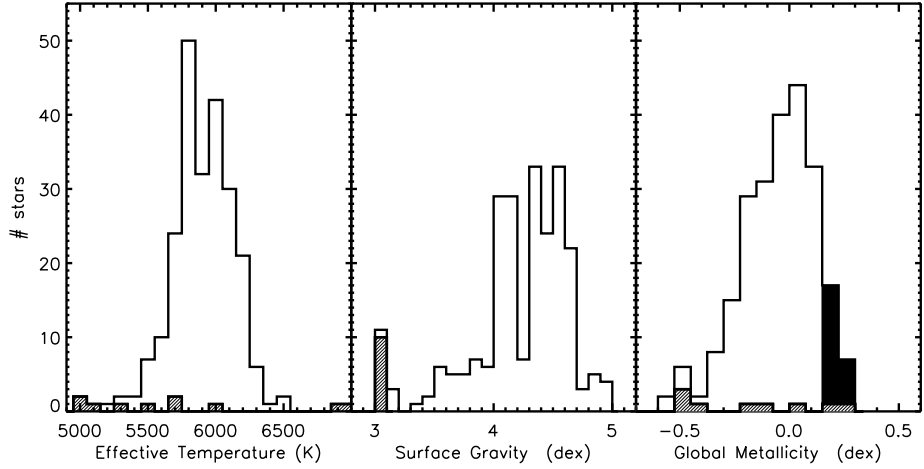}
\caption{Atmospheric parameters distributions of the sample of target stars. The shaded area shows the locations of the outliers (see text). The black area shows the SMR stellar sample. Fot the sake of clarity, in the right panel we do not include the ``outlier'' BD+34~4028, for which we determined [M/H]=$-$2.1.}
\label{fig:distr}
\end{figure*}

\begin{enumerate}
\item for each $i$-th index, we computed the average value of a moving standard deviation, with a 5~\AA-wide window, over the whole index wavelength interval (from the bluer to the redder limit) on the observed spectrum, and we assumed this value as the estimation of the noise level. Note that, during the first iteration, the noise is definitely overestimated, because the observed spectrum contains all the absorption lines, whose contribution is minimized by using a moving standard deviation, but not completely eliminated;
\item  the computed noise level is assumed as the standard deviation of a Gaussian distribution which is used to randomly add noise to each wavelength point inside the index interval, and then the index is calculated. This step is repeated 1000 times for each index, and we end up with distributions of index values, whose standard deviations provides $\sigma^{2}_{i,obs}$, which are entered in Eq.~\ref{eq:chi2} to estimate the stellar parameters;
\item these parameters are interpolated in the Munari's library and the 
corresponding spectrum is subtracted to the observed one to obtain a residual 
spectrum. In the ideal case, this spectrum should contain only the observational noise, but, since the synthetic spectra do not perfectly match observations, we can still expect some trace of spectral features. This is well described in Fig~\ref{fig:resid}, where we show the residual spectrum, after the first iteration, for the index Fe4531 of the star BD+60~600;
\item the process is repeated until two consecutive iterations produce the same set of atmospheric parameters.
\end{enumerate}

We used the distribution of $\chi^2$ values to estimate the error on the parameters, 
following \citet{avni76}. This is, all the combinations of parameters that have $\chi^2<{\rm 
MIN}(\chi^2)+3.5$ generate a volume in the parameters space, whose extreme values 
along the three dimensions give the (commonly asymmetric) 68\% confidence level error.
The results are reported  in Table~\ref{tab:param} where we list the set of stellar parameters 
and the corresponding error estimates.

\section{Results and comparison with previous works}

\begin{figure}
\includegraphics[width= 84mm]{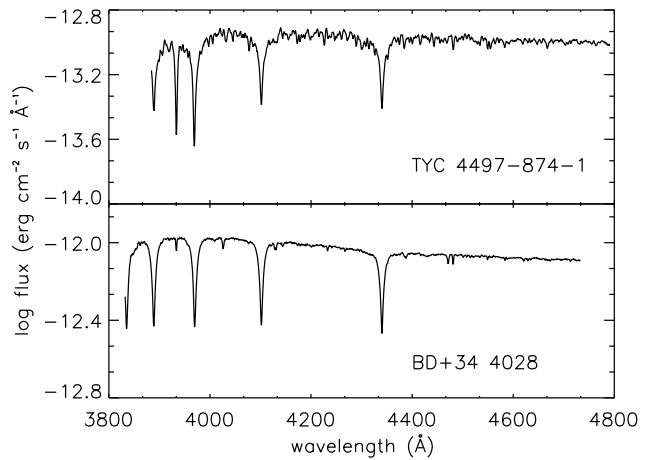}
\caption{Spectra of the two stars with the highest $T_{\rm eff}$.}
\label{fhotstar}
\end{figure}

In Fig.~\ref{fig:distr} we show the distributions of the three atmospheric parameters for the 
233 stars in our sample which in general appear compatible with those expected for G0--G3 stars 
on the main sequence.
Note that the dispersion of the surface gravity distribution is quite large which, 
as previously pointed out, might be ascribed to the lower precision of 
the adopted method in determining this parameter. The average 1$\sigma$ error associated to $\log{g}$ is 0.26 dex, 
while it is 0.056~dex for [M/H] and 55~K for $T_{\rm eff}$. These figures are obtained once the outliers, indicated 
with a shaded area in
Fig.~\ref{fig:distr}, are removed. We have identified as outliers the 10 objects for which the best fit described in 
the previous section yields a surface gravity at the lower boundary of the synthetic grid ($\log{g}=3.0)$.
Two of these outliers, TYC~4497-874-1 and BD+34~4028, have $T_{\rm eff}$ much larger than the bulk of the sample (7000~K and 6890~K, 
respectively), and in fact their spectra, depicted in 
Fig.~\ref{fhotstar}, clearly indicates that they are F or late-A stars. Their $B-V$ colors are compatible with higher temperature objects: from SIMBAD these colors are 0.45 for TYC~4497-874-1
and 0.23 for BD+34~4028, which according to the color index-spectral calibration of \citet{pecaut12}, and assuming no 
extinction, would correspond to stars of types  F5--F6 and A7--A8, respectively. The spectral classification of these
stars, in particular that of BD+34~4028, which dates back to the early sixties \citep{barbier62}, should be revisited.

It remains to be investigated the origin of the discrepancies for the rest of the deviant objects. Plausible 
explanations can be, among others,  that we are actually dealing with binaries with composite spectra 
that cannot be separated at the working resolution, objects that are highly variable or that (some of) 
these objects are also wrongly classified. The answer is beyond the scope of this paper. For now, the 
parameters of the outliers should be considered as uncertain, in particular for the other 8 stars with 
$\log{g}=3.0$~dex: BD+31~3699,  BD+42~384, BD+58~681,  BD+60~2506, HD~105898, HD~149996, 
BD+45~2871 and TYC~3619-1400-1. 
The 10 ``outliers'' can be easily identified in
the electronic version of Table~\ref{tab:param} and are indicated with the symbol "*" after the stellar identification.

\begin{figure*}
\includegraphics[width = 148mm]{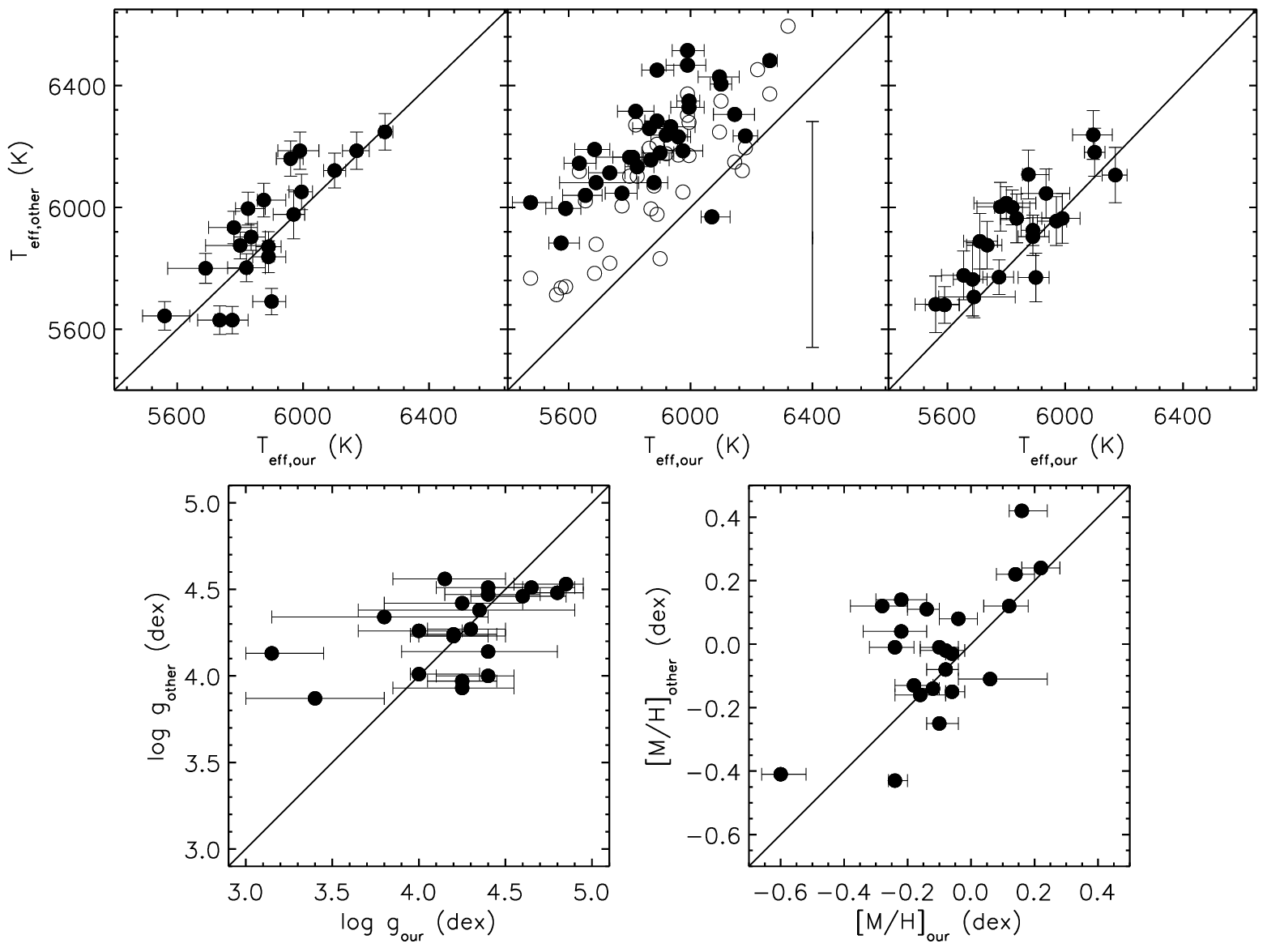}
\caption{{\it Top panels:} Comparison of the $T_{\rm eff}$ determined in this work and the determinations by 
\citet{masana06} (left), \citet{bailer11} (middle) and \citet{casagrande11} (right). The solid line indicates the slope unity. In the middle panel, the filled and empty circles stand, respectively, for the results of the p and pq models of \citet{bailer11}. For the sake of clarity, we have incorporated in this panel a bar indicating the average value of the 90\% confidence intervals reported in \citet{bailer11}. {\it Bottom panels:} Comparison of the $\log{g}$ (left) and [M/H] (right) reported in \citet{casagrande11} and those derived in this study.}
\label{comp}
\end{figure*}

\subsection{Comparison with previous works}
\label{sec:comp}
In addition to the test with the solar spectra, a natural exercise to verify the global 
validity of the set of parameters we have derived is to extend the comparison with other determinations from the literature: we selected three sources that contain several of our target stars.

In the top panels of Fig.~\ref{comp}, we illustrate the comparison of the  $T_{\rm eff}$ for 
the 20 stars in common with \citet{masana06} (left panel), who determined this parameter from $V$ and 2MASS infrared photometry, with the study of \citet{bailer11} (middle panel), who also provides $BVJHK$ photometric estimation of $T_{\rm eff}$  for the objects in common (34 with his {\em p-model} and 37 with his {\em pq-model}), and with the work of \citet{casagrande11} (right panel), which also includes 22 stars of our sample and obtains $T_{\rm eff}$ by means of the infrared flux method \citep{casagrande06}.
By inspection of these figures, we can readily see that our
resulting effective temperatures show, albeit some dispersion, an overall good agreement with 
\citet{masana06} and \citet{casagrande11}, while are highly discrepant, of the order 
of 250~K lower, when compared with the {\em p-model} of \citet{bailer11}. The disagreement, cannot be explained by the fixed gravity and metallicity he 
used in his calculations. \citet{bailer11} data are consistent with the temperature scale of \citet{bailer97} that indicates
a temperature for a G2V star of 6015$\pm$49~K, which is more than 200~K higher than the accepted value 
for the Sun. Significant differences have been also found by \citet{waite11} for the star HIP 68328: these 
authors find a temperature lower by about 450~K than that of \citet{bailer11}, which somewhat agrees with the average difference with our results at low temperature. 
The comparison with the {\em pq-model} temperatures of \citet{bailer11} shows a slightly smaller, but still systematic, offset.

For the comparison of $\log{g}$ and [M/H], we have used the stars in common 
with \citet{casagrande11}. They derive [Fe/H] from a calibration of Str\"omgren colors and $\log{g}$ from the fundamental relation involving mass, $T_{\rm eff}$, and bolometric luminosity. 
The comparison is depicted in the lower panels of Fig.~\ref{comp} for the 22 stars 
that have both parameters available. The agreement is, in general, good. We have already 
mentioned that our gravity determinations are prone to large uncertainties, which translate in a larger dispersion of our data. However, our average $\log{g}$ of 4.23~dex is consistent with the 4.27~dex value of \citet{casagrande11}. Regarding 
metallicity, it appears that we somewhat underestimate it, as the average value of the global abundances for
\citet{casagrande11} data and ours are, respectively, -0.02 and -0.10~dex. Nevertheless, \citet{casagrande11} state that their metallicity scale is 0.1~dex higher than the previous results they use for comparison.

\subsection{SMR stars}
According to the results presented in Sect.~\ref{sec:det}, we identify 22 SMR stars, of which 20 are new identifications, with a global 
metallicity in excess of $+$0.16~dex, the most metallic being TYC 2655-3677-1, with [M/H]=$+$0.28$^{+0.04}_{-0.06}$. 
This sample represents about 10\% of the observed targets and about the same proportion when compared with 
the set of stars whose average metallicity exceeds the above value in the PASTEL catalogue. 

\begin{figure}
\includegraphics[width = 84mm]{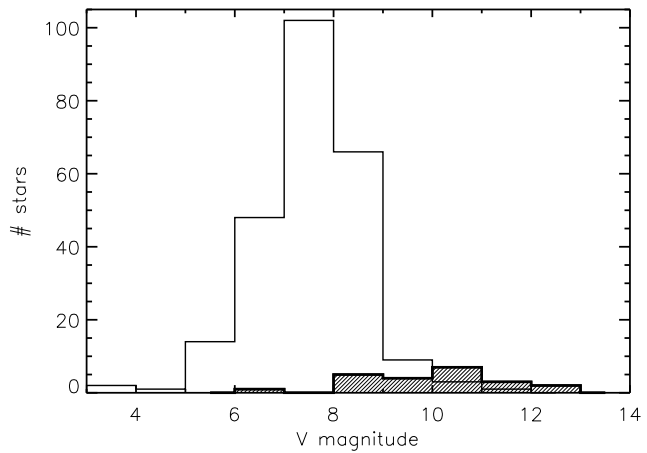}
\caption{$V$ magnitude distribution of SMR stars of this work (shaded area) and from PASTEL catalogue.}
\label{f11}
\end{figure}

In Fig.~\ref{f11}, we depict the $V$ band distribution of the sample 
of SMR stars (shaded histogram) compared to the distribution of the 246 SMR stars 
present in the PASTEL catalogue. This latter sample was selected based upon
these criteria: $5400<T_{\rm eff}<6400$~K, $4.0<\log{g}< 5.0$~dex and [M/H]$\geq$~$+$0.16~dex. 
The number of SMR stars with $V>8$ is significantly increased, from 79 to 100, as a result of our study. 
The stellar parameters of SMR objects 
are given in Table~\ref{tab:smr}.

\begin{table*}
\begin{minipage}{126mm}
\caption{Atmospheric parameters of SMR stars.} 
\label{tab:smr}
\begin{tabular}{lcccccccccc}\hline
{Object} & {T$_{\tiny \mbox{eff}}$ (K) }& \multicolumn{2}{c}{$\sigma_{\tiny \mbox{T}_{\mbox{eff}}}$ (K)} &{$\log g$ 
(dex)} & \multicolumn{2}{c}{$\sigma_{\tiny \mbox{log g}}$ (dex)} & {$\left[\mbox{M/H}\right]$ (dex)} & 
\multicolumn{2}{c}{$\sigma_{\tiny \mbox{[M/H]}}$ (dex)} \\
   & & + & - & & + & - & & + & - \\ \hline 

TYC 1759-462-1              &   6140   &     80   &     75   & 3.65   & 0.45   & 0.40   & 0.20   & 0.12   & 0.12\\ 
BD+60 402                   &   5985   &     75   &     70   & 4.30   & 0.40   & 0.40   & 0.22   & 0.08   & 0.10   \\
TYC 1230-576-1              &   5665   &     85   &     75   & 4.40   & 0.35   & 0.30   & 0.16   & 0.10   & 0.08   \\
HD 232824                   &   5900   &     50   &     85   & 4.15   & 0.25   & 0.45   & 0.16   & 0.06   & 0.10   \\
HD 237200                   &   6045   &     40   &     70   & 4.25   & 0.25   & 0.40   & 0.18   & 0.04   & 0.06   \\
HD 135633                   &   6095   &     65   &     70   & 4.25   & 0.35   & 0.45   & 0.22   & 0.06   & 0.06   \\
BD+28 3198                  &   5840   &     25   &     45   & 4.00   & 0.10   & 0.25   & 0.24   & 0.04   & 0.06   \\
Cl* NGC 6779 CB 471         &   5790   &     70   &     65   & 3.80   & 0.35   & 0.35   & 0.20   & 0.10   & 0.10   \\
TYC 2655-3677-1             &   6220   &     45   &     50   & 4.15   & 0.30   & 0.25   & 0.28   & 0.04   & 0.06   \\ 
HD 228356                   &   6055   &     55   &     20   & 4.00   & 0.30   & 0.10   & 0.16   & 0.06   & 0.04   \\
BD+47 3218                  &   6050   &     45   &     60   & 4.05   & 0.25   & 0.35   & 0.16   & 0.04   & 0.08   \\
\rm{[M96a]} SS Cyg star 14  &   6195   &     35   &     40   & 4.00   & 0.25   & 0.25   & 0.24   & 0.04   & 0.06  \\
TYC 3973-1584-1             &   6000   &     35   &     55   & 4.45   & 0.20   & 0.30   & 0.20   & 0.04   & 0.06  \\
BD+52 3145                  &   6095   &     35   &     50   & 3.95   & 0.20   & 0.25   & 0.26   & 0.06   & 0.06  \\
TYC 3986-3381-1             &   5855   &     55   &     60   & 4.15   & 0.25   & 0.25   & 0.26   & 0.08   & 0.06  \\
TYC 3982-2812-1             &   5895   &     60   &     50   & 4.30   & 0.30   & 0.25   & 0.18   & 0.06   & 0.06  \\
TYC 3618-1191-1             &   5940   &     60   &     60   & 4.40   & 0.35   & 0.30   & 0.24   & 0.08   & 0.08 \\
TYC 3986-758-1              &   5845   &     55   &     65   & 4.05   & 0.30   & 0.35   & 0.16   & 0.06   & 0.08\\         
BD+60 600                   &   5655   &     35   &     60   & 3.95   & 0.10   & 0.30   & 0.20   & 0.06   & 0.08  \\
HD 283538                   &   6005   &     25   &     40   & 4.00   & 0.10   & 0.25   & 0.16   & 0.04   & 0.06 \\
HD 137510                   &   5875   &     70   &     20   & 4.00   & 0.35   & 0.05   & 0.16   & 0.08   & 0.04   \\
HD 212809                   &   5975   &     60   &     50   & 4.55   & 0.30   & 0.25   & 0.16   & 0.04   & 0.06   \\
\hline
\end{tabular}
\end{minipage}
\end{table*}

\section{Summary}
The stellar parameters (T$_{\rm eff}$, $\log g$, \rm{[M/H]}) 
of a sample of 233 stars with spectral types
between G0V-G3V were determined through the comparison of 
a set of spectroscopic indices and those calculated from a segment of Munari's library of
 synthetic spectra. This work provides 
the first spectroscopic determination of the stellar parameters for almost 70$\%$ (213) 
of the stars in our sample. The results obtained are compatible
with the expected values for the considered spectral types, and in general 
agreement with previous determinations. In particular the effective temperature
of \citet{casagrande11} and \citet{masana06}. We nevertheless find significant
incosistencies with \citet{bailer11}, particularly at the lower temperature 
edge ($\sim$5600~K).

We identified a new sample of 20 SMR stars plus two that were already known. 
The comparisons presented in Sect.~\ref{sec:comp} and that conducted with the solar spectrum 
gives us confidence that the SMR stars found in this work indeed represent 
bona fide targets for future searches of giant exoplanets. Additionally, the present sample
and its planned extensions in number of objects and in wavelength converage will complement, for instance,
the data set of \citet{adibekyan11} and the fainter sample
included in \citet{lee08} in the investigation of the
chemical evolution of the Galaxy.

\section*{Acknowledgments}
R.L.V, E.B. and M.C. want to thank CONACyT for financial support through grants
SEP-2009-134985 and SEP-2011-169554. This research has made use of 
the SIMBAD database, operated at CDS, Strasbourg, France.

\bsp

\label{lastpage}

\end{document}